\documentclass{sig-alternate-10pt}
\usepackage{epsfig,endnotes}
\usepackage{color}
\usepackage{amsmath,amssymb}
\usepackage[scriptsize,tight,center]{subfigure}
\usepackage[normalem]{ulem}
\usepackage{algorithmic}
\usepackage{algorithm}
\usepackage{enumitem}
\usepackage{soul}
\usepackage{comment}
\newenvironment{Itemize}%
{\begin{itemize}%
\setlength{\itemsep}{0pt}%
\setlength{\topsep}{0pt}%
\setlength{\partopsep}{0pt}%
\setlength{\parskip}{0pt}}%
{\end{itemize}}

\newcommand{\bcomment}{\color{blue}}

\begin{document}

%\title{Helping Alice connect:\\
%adapting connectivity over physical layer cooperation networks}
%a scheme for physical layer cooperation approach}

%\title{Which relays to use:\\ keeping the connection strong
%adapting connectivity over physical layer cooperation networks}
% keeping connectivity strong, picky, discriminating

%\author{
%{\rm Ayan Sengupta}\\
%UCLA, USA
%\and
%{\rm Yahya H.\ Ezzeldin}\\
%UCLA, USA
%\and
%{\rm Siddhartha Brahma}\\
%University of Neuchatel, Switzerland
%\and
%{\rm Christina Fragouli}\\
%UCLA, USA
%\and
%{\rm Suhas Diggavi}\\
%UCLA, USA
%} % end author

\author{
{Ayan Sengupta$^\star$\qquad Yahya H. Ezzeldin$^\dagger$\qquad Siddhartha Brahma$^\ddagger$} \\ {Christina Fragouli$^\dagger$\qquad Suhas Diggavi$^\dagger$}\\\\
    $^\star$Stanford University, USA \qquad $^\dagger$UCLA, USA\qquad  $^\ddagger$IBM Research - Almaden, USA
} % end author
\title{Consistency in the face of change:\\ an adaptive approach
 to physical layer cooperation }
 \maketitle

\begin{abstract}
Most existing works on physical-layer (PHY) cooperation (beyond routing) focus on how to best use a given, static relay network--while wireless networks are anything but static.  In this paper, we pose a different set of questions: given that we have multiple devices within range, which relay(s) do we use for PHY cooperation, to maintain a consistent target performance? How can we efficiently adapt, as network conditions change? And how important is it, in terms of performance, to adapt? Although adapting to the best path when routing is a well understood problem, how to do so over PHY cooperation networks is an open question. Our contributions are: (1) We demonstrate via theoretical evaluation, a diminishing returns trend as the number of deployed relays increases. (2) Using a simple algorithm based on network metrics, we efficiently select the sub-network to use at any given time to maintain a target reliability. (3) When streaming video from Netflix, we experimentally show (using measurements from a WARP radio testbed employing DIQIF relaying) that our adaptive PHY cooperation scheme provides a throughput gain of 2x over nonadaptive PHY schemes, and a gain of 6x over genie-aided IP-level adaptive routing.
\end{abstract}

\section{Introduction}
\label{sec:introduction}
Physical layer (PHY) cooperation can help relieve the bandwidth crunch that
remains a core threat to mobile user experience today. %In the information theory and communications literature,
A number of theoretical works have established the significant
benefits enabled by PHY cooperation; system implementations
have also validated that it is feasible for current devices to support
PHY cooperation  \cite{MobihocQMF,Quilt2014,80211ad}.
However, a crucial missing piece remains:  how does one implement PHY adaptation over cooperative networks?
Most works {beyond routing} have focused on optimally operating a given static relay, while today, in our
homes, at work and in public places, we can have changing network topologies
and  multiple wireless (\emph{e.g.,} WiFi-enabled) devices within range to assist a source-destination pair.

Consider for instance the following scenario: Alice is streaming Netflix to her tablet using WiFi, and takes the tablet with her as she moves around her house\footnote{{Our application focus is on indoor scenarios, where we do not expect ``fast-fading'', as can occur in (LTE-based) cellular networks for objects moving at high velocities.}}. Her house has a WiFi router wired to the Internet that acts as source, as well as three other wireless devices that can act as WiFi relays, and can implement PHY cooperation with the source. As her network conditions change, which (if any) of the three potential relays should assist her, so that she can experience a consistently good video quality? This is a well studied problem in routing, where one selects which relays to route packets through; in this paper we focus on PHY cooperation networks, {where there is (potentially) a larger gain over adaptive routing, yet practical algorithms for exploiting the same are less well understood.}

We are interested in the following questions: Should we select one or more relays to assist us via PHY cooperation? What are the best relay(s) to use? Is it easy to identify them? How can we efficiently adapt our choice as Alice moves? How important is it, in terms of performance, to adapt? And what are the benefits of adaptive PHY cooperation over adaptive routing? Answering such questions is a necessary step in bringing PHY cooperation closer to practical systems.

We restrict our attention to using either none, one or two relays, and  at most two cooperating transmitters at a time. We make this choice for two reasons. First, we do not expect significant benefits from simultaneously using three or more relays; this is supported by numerical evaluations of outage performance (in Section~\ref{sec:theoretical_evidence}) that indicate a ``diminishing returns'' trend for larger number of relays--the more relays we use, the less we gain per relay. Second, the complexity of the system increases exponentially with the number of simultaneously transmitting nodes, as we discuss in Section~\ref{sec2.4}.

{We build upon our PHY cooperation scheme DIQIF \cite{Quilt2014} and customize it for two relay cooperation. We explore the benefits that two relays can enable in practice.}

%We find that in some configurations, engaging two relays  can improve the Frame Error Rate (FER) performance by almost two orders of magnitude. However, in other scenarios, we get significantly better performance (an order or magnitude better) by engaging only one of the two relays (in cooperation with the source) in the cooperative phase. Hence, it is not the case that engaging two relays is consistently better than engaging one, when using two-transmitter PHY cooperation.

 %; unlike setups such as MIMO, where we can perform beamforming and successively decode multiple superimposed signals, we here look at the case of a single antenna receiving a superposition of signals that need to be simultaneously processed for decoding, as dictated by the information-theoretic paradigm of Quantize-Map-and-Forward \cite{ADT11}, which ideologically inspired the DIQIF strategy \cite{Quilt2014}. As a result, the complexity of the decoders increases exponentially with the number of superposed signals at the destination; in our experimental work, we used customized belief propagation decoders that are the state-of-the-art in joint decoding at low complexity.

{Next, we} introduce SPA, an  algorithm that selects whether to use a subset of one or two relays, and exactly which subset, from among a set of  available relays, by performing Frame Error Rate (FER) learning. Our contribution is a design that carefully balances the training overhead with the estimation performance of SPA, and gracefully adapts to channel changes. Inspired from machine-learning techniques, SPA learns the best subset while sending the actual data we want to transmit (as opposed to dummy training data), invokes an early reject criterion while learning to quickly weed out relays that are not likely to lead to good performance, and uses memory to adapt to changing network conditions. In addition, SPA is completely agnostic to the underlying PHY relaying scheme, and can be used in conjunction with other strategies like Decode-Forward, Amplify-Forward etc., if desired.

Our next contribution is a proof-of-concept deployment on a WARP software radio testbed. We find that DIQIF outperforms alternative schemes for two-relay cooperation; that performance is significantly affected by the choice of cooperating relay(s); that SPA can reliably select an operational mode that enables consistently low FERs in the face of changing network conditions; and that the benefits from adaptation in SPA can enable Netflix streaming at higher (and consistent) rates, promising a better user experience. Interestingly, when pitted against protocols without PHY cooperation, SPA significantly outperforms ($\geq 6$x Netflix throughput gain) even a genie-aided adaptive routing strategy with knowledge of future network conditions.

In what follows, Section~\ref{sec:related_work} reviews related work; Section~\ref{sec:cooperation_model} presents the relaying scheme and theoretical evidence through numerical evaluation; Section~\ref{sec:adaptive_algorithms} introduces SPA, our algorithm for relay selection; Section~\ref{sec:system_implementation} describes system implementation; Section~\ref{sec:experimental_results} presents experimental evaluations; Section~\ref{sec:video_evaluation} tests video streaming performance, and Section~\ref{sec:conclusions} concludes the paper.

\section{Related Work}\label{sec:related_work}
%\noindent The related work falls into three categories:\\
\noindent{\bf Relay Selection Algorithms:}
\iffalse\textcolor{red}{Cant we say anything more apart from that they do not offer testbed evaluations? Any reason why we think they would not work so well as SPA?  the reviewers could say, implement them and compare against them. Also: we create the impression that SPA would only work with DIQIF - not true!}\fi
{ Theoretical algorithms (of both distributed as well as centralized
nature) for relay selection are studied in \cite{Guan2011, Wang2009, Shi2008, Jing2009, LoHeath2009} and the references therein.
In \cite{Guan2011,Wang2009,Jing2009,Shi2008} orthogonalized single-carrier transmissions, as well as Amplify-Forward (AF) and/or Decode-Forward (DF) relaying strategies are considered and algorithms are proposed that use an objective function based on channel state information (CSI).
In \cite{LoHeath2009} a version of the Partial DF \cite{Erkip2005} relaying scheme is studied and algorithms based on CSI as well as distance are proposed. However, the objective functions for relay selection in \cite{LoHeath2009} are based on high-SNR approximations. These theoretical works do not offer testbed evaluations of their algorithms in real-world scenarios. Moreover, the schemes considered--AF and DF, have also been shown to perform suboptimally in theory \cite{ADT11} as well as practice \cite{MobihocQMF}. Additionally, and differently from our approach, \cite{LoHeath2009, Jing2009} assume accurate forward CSI at the transmitter which can be difficult to acquire in wireless systems and also represent an overhead bottleneck in multicarrier (OFDM-based) systems, while introducing errors in SNR-limited regimes.
%In our testbed evaluation, using DIQIF, we found that the ordering based on the objective functions in the above work did not necessarily provide an accurate ordering of the performance of the selected relays in terms of Frame Error Rate (FER).
%{\color{blue} Dynamic DF scheme \cite{Azarian2005}, is another candidate scheme that is optimal at high SNR regimes but doesn't improve the outage at low SNR in comparison to DF \cite{Avestimehr2007}.}
%Furthermore, it is not understood how to extend its capabilities to multiple relays.
Multiuser scheduling algorithms for
exploiting cooperation benefits from OFDM-based DF relay
networks are studied in \cite{Deb2008}. Hop-by-hop best relay
selection techniques for exploiting diversity in routing-based
multihop wireless networks are explored in \cite{Jain2008}.}\\

\noindent{\bf Network Information Theory:}
Our work is inspired by the information-theoretic result on wireless network simplification \cite{ayfersimplify},
which shows that using subsets of relays can achieve a constant fraction of the capacity;
\cite{ayfersimplify} provides capacity results for static Gaussian channels,
while we focus on adaptation over time-varying configurations and practically implementable relay selection protocols. \cite{ADT11} presents the QMF scheme that approximately achieves the network capacity and which is the foundation for the DIQIF scheme \cite{MobihocQMF} used in this paper.\\

\noindent{\bf PHY Cooperation Testbeds:}
For concurrent sender-receiver pairs, Analog Network Coding and interference alignment based system implementations were proposed in \cite{Katti07} and \cite{Gollakota2009}, respectively.
The first testbed implementation of QMF over single-relay
systems, as well as that of coded non-orthogonal AF and DF relaying
systems were presented in \cite{MobihocQMF}, demonstrating significant benefits of QMF over AF and DF. \cite{Quilt2014} extended the work
of \cite{MobihocQMF} to include opportunistic
decoding/quantization at the relay and hybrid decoding at the
destination. Other WARP \cite{Khattab2008} based implementations of
PHY cooperation networks include the work of
\cite{RiceWork}, where uncoded AF and DF relaying were used.
MAC-PHY layer protocols for optimally
triggering cooperative transmissions in WiFi networks
were studied in \cite{Bejarano2013,SourceSync,AirSync}.

\section{Cooperative Transmission Scheme}
\label{sec:cooperation_model}
%====================================
We here first describe the network operation and give an overview of DIQIF \cite{Quilt2014}. Our new contributions are that we customize DIQIF to $2$ relay networks, and present theoretical evidence to support our design choices.

\subsection{Network Operation}
\paragraph*{Modes} Consider a source S that wants to communicate with a
destination D.
We have a direct link
%\footnote{The direct link may either be strong or so weak that it is essentially nonexistent, depending on the configuration.}
connecting S to D, as
well as $N$ relays available to help the S--D
pair. Our goal is to adaptively select which 1 or 2
relays, among the $N$, will assist the S--D communication if needed, at each point in time.
Clearly, we have $N$+ ${N}\choose{2}$ choices;
we term these choices ``modes''. We have $N$ ``1-relay modes''  where the source  cooperates with a specific
relay $i$, and ${N}\choose{2}$ ``2-relay modes'' where the source cooperates with two specific relays $i$ and $j$.
\begin{figure}[t]
  \centering
  \includegraphics[width=0.65\columnwidth]{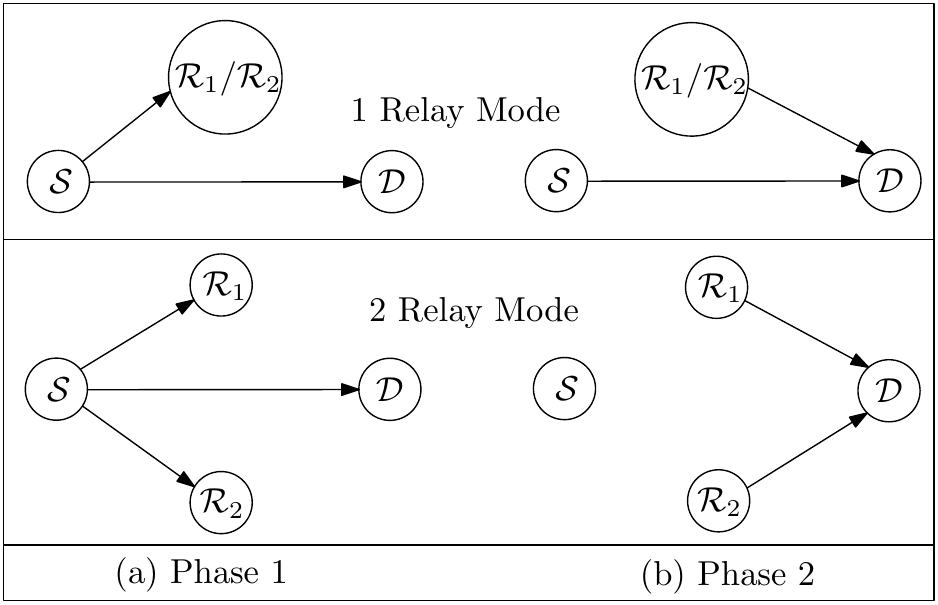}
  \caption{Two-Stage Network Operation}\label{fig:two_phase_operation}
  \vspace{-1em}
\end{figure}
\vspace{-2mm}
\paragraph*{Two phase operation}
In both types of modes, the source first attempts to communicate with the destination; if it fails,
a cooperative transmission takes place: \\
$\bullet$ {\em 1-relay mode.} In Phase $1$ the source transmits; the
  destination and the {one} relay in the mode listen to the transmission. If the
  destination cannot decode, in Phase $2$, the source and the
  relay cooperatively transmit; the destination receives the
  superposition of the source and the relay signals.\\
$\bullet$ {\em 2-relay mode.} In Phase $1$ the source S transmits; the
  destination D and the {two} relays  listen to the transmission. If the
  destination cannot decode, in Phase $2$, the two relays
  cooperatively transmit; D receives the superposition
  of the two relay signals. %\end{Itemize}

Fig.~\ref{fig:two_phase_operation} depicts, for $N=2$,
the three possible modes and the two-phase operation, where one or both relays,
respectively, assist the S--D communication. Since we restrict our attention to two cooperative transmitters at any time,
we do not consider the case where the source would transmit simultaneously with the two relays; we discuss why in Section~\ref{sec:theoretical_evidence} (P4) and Section~\ref{sec2.4}.

\subsection{Relaying Strategy}
\begin{figure*}[!t]
  \centering
\subfigure[Clustered Network.]
  {\includegraphics[width=0.63\columnwidth]{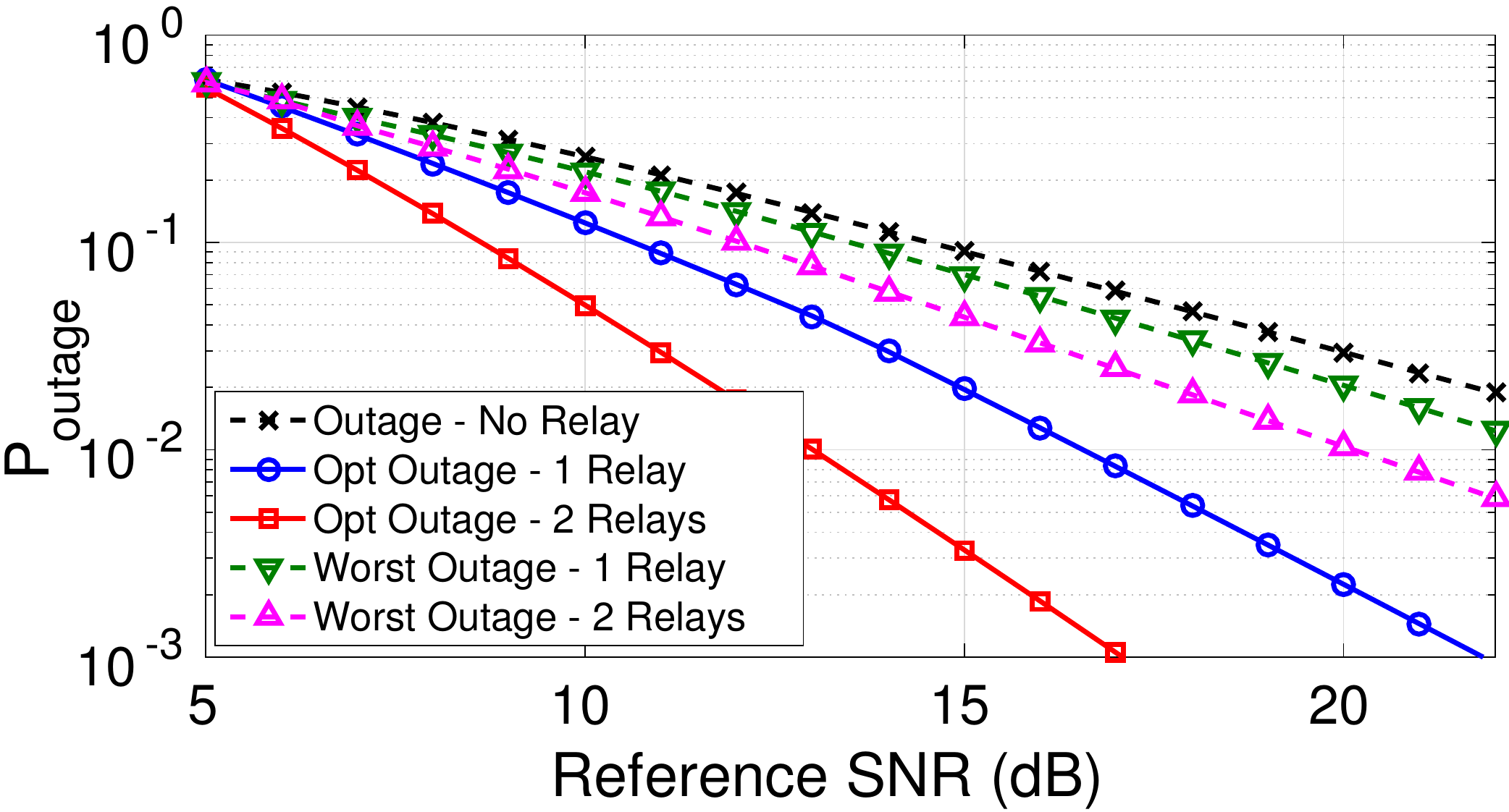}
\label{fig:outage_progression__bestworst_clus}
}
\subfigure[Clustered Network.]
  {\includegraphics[width=0.63\columnwidth]{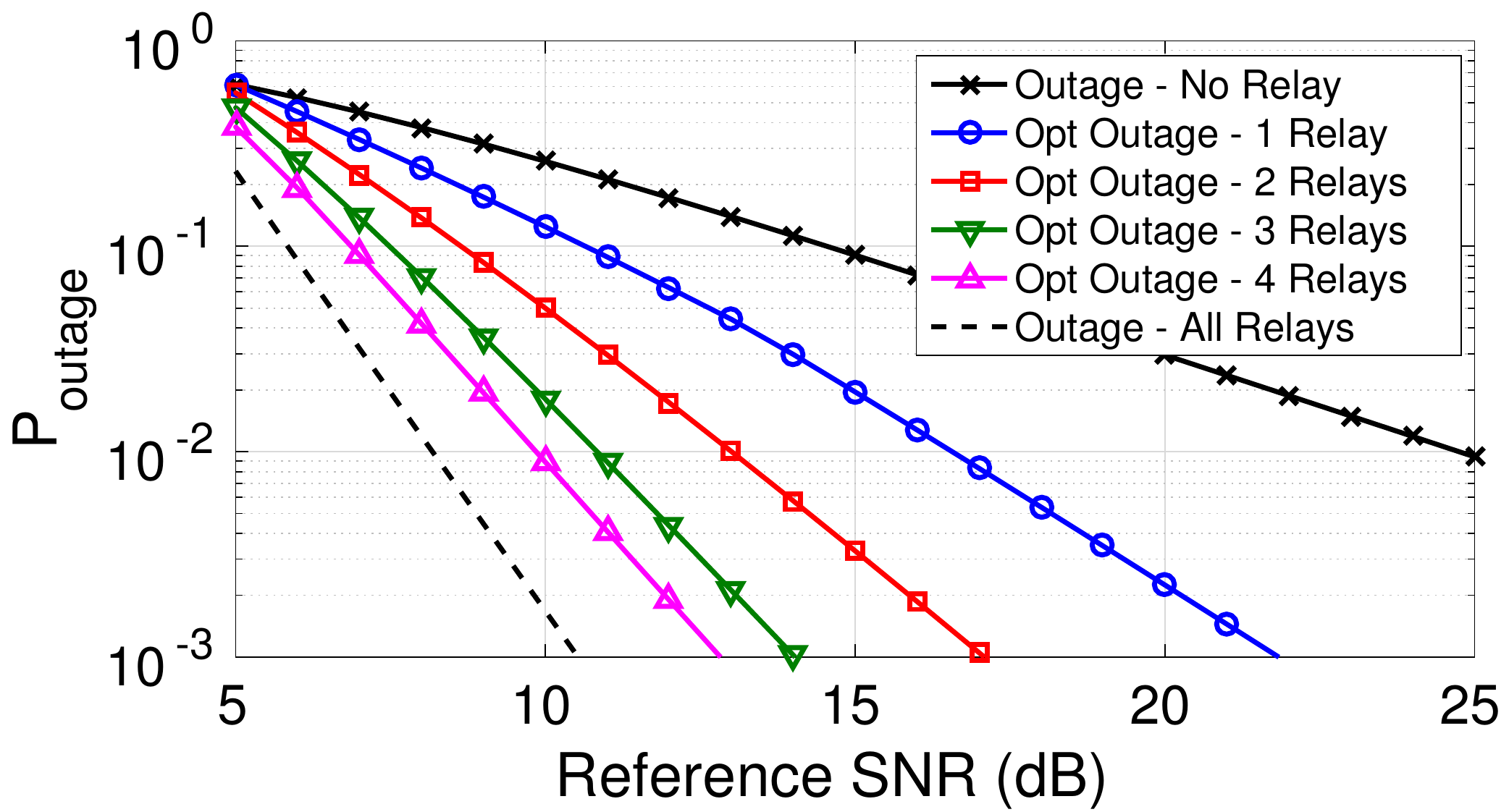}
\label{fig:outage_progression_clus}
}
\subfigure[Symmetric Network.]{
  \includegraphics[width=0.63\columnwidth]{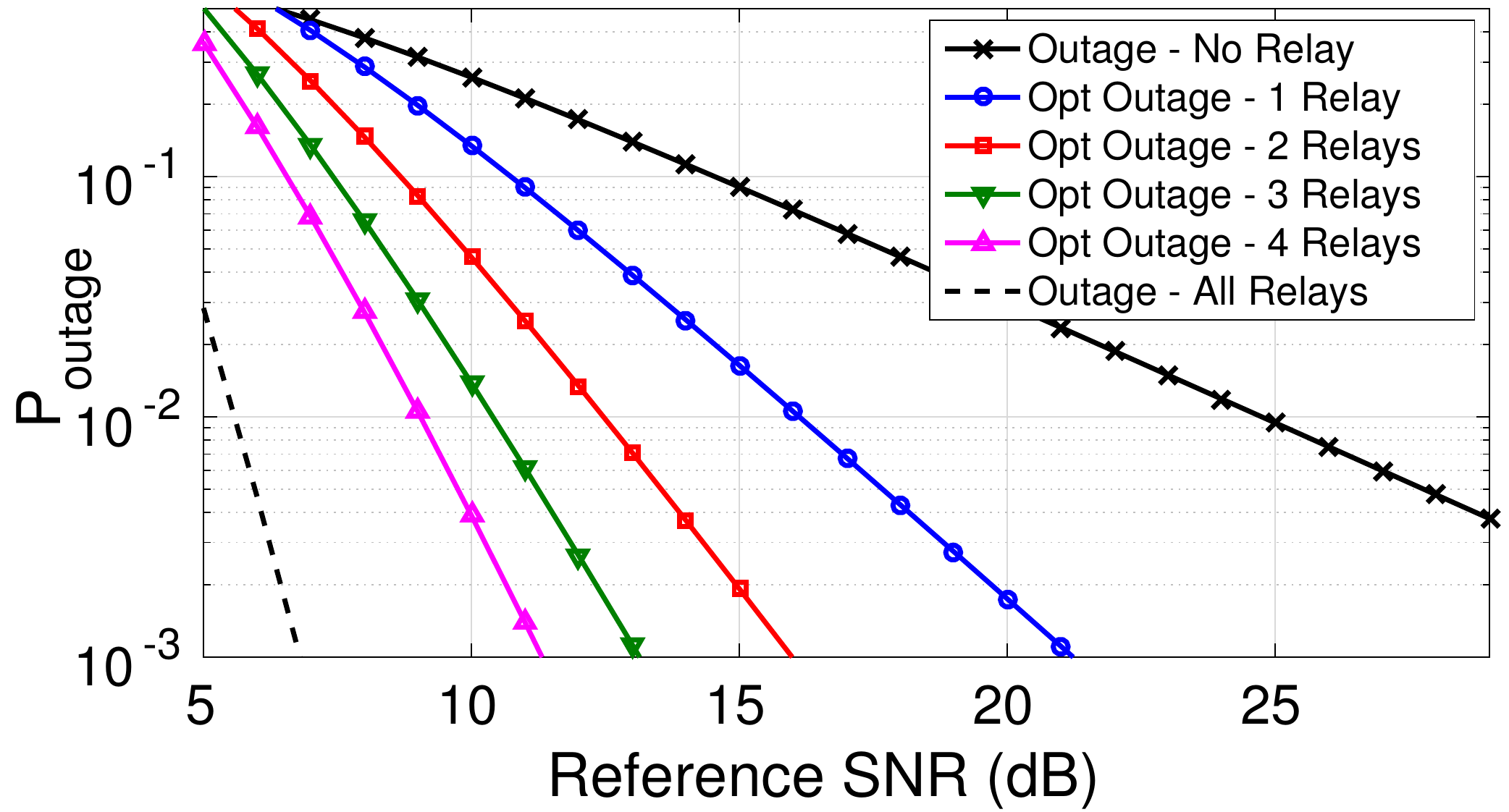}
\label{fig:outage_progression_iid}
}
\caption{Outage performance over Rayleigh-faded relay
    networks with increasing number of relays selected out
    of maximum of $10$ relays. The $x$-axis denotes \emph{per node} reference SNR.}
\label{fig:outage_prog_overall}
\vspace{-1em}
\end{figure*}
%We use the relaying operation proposed in our prior work, DIQIF \cite{Quilt2014}, customized for 2-relay networks.
%Following \cite{Quilt2014},
We implement DIQIF relaying over the PHY procedures of WiFi, such as  OFDM modulation. The source transmission is always a coded information stream using standard (for WiFi) LDPC codes.

DIQIF \cite{Quilt2014} for a $1$-relay mode operates as follows. The relay first attempts to decode its received signal.
If it succeeds, then the relay has access to the clean information bits. It re-encodes the information bits with the same code as the source, and interleaves the coded bits. If it does not succeed in decoding, the relay quantizes the elements of its received signal
to their closest constellation points, and interleaves the resulting bits. The relay and the source then synchronously transmit the coded sequence (after appropriate OFDM modulation) in Phase II (if Phase I fails), effectively creating a distributed space-frequency code.

To extend this scheme to two relays, we apply the same operation (if possible decode, otherwise quantize) at each relay, and have the two relays synchronously transmit in Phase~II. However, we ensure that each relay uses a {\em different} interleaver, as we found that this is  crucial in achieving diversity, and thus a good error performance.
% We used randomly selected interleavers at each relay, and found that using different interleavers is crucial in achieving diversity, and thus a good error performance; however, which %{specific} interleaver each relay used, hardly affected performance.
 {For decoding at the destination, we designed a belief-propagation joint decoder, that takes into account the structure of the (two-relay) network (including quantizers at the relays, and superposition at the destination), and jointly decodes from the (different) relay transmissions of the source message.}
\iffalse
\begin{figure}
  \centering
  \includegraphics[width=0.8\columnwidth]{paper_fig_decoder}\\
  \caption{LDPC-based joint decoder for one codeword of DIQIF. The quantizer nodes in the graph are replaced by deterministic connections for codewords where relay decoding succeeds.}\label{fig:QMF_decoder}
  \vspace{-1em}
\end{figure}
\fi
\subsection{Theoretical evidence}\label{sec:theoretical_evidence}
We here provide evidence, through theoretical modeling and simulations,
to support the following points:
\begin{Itemize}
\item[\em P1.] There can be a significant performance variability depending on which relay(s) we select.
\item[\em P2.] As long as we select a few of the relays carefully, we can already extract a significant portion of the cooperation benefits that the full network offers.
\item[\em P3.] As the number of relays increase, we get diminishing returns in terms of  additional benefits.
{\item[\em P4.] In a $2$-relay network, having both relays and the source simultaneously transmit does not offer significant benefits over the best 2-transmitter mode.}
%The best $2$-transmitter mode in a $2$-relay network incurs only minimal loss with respect to the $3$-transmitter bound--where %the source and both the relays jointly transmit to the destination.
\end{Itemize}
We emphasize that this evidence is only {indicative} of the trends we expect in our testbed: analyzing the  performance of cooperative networks that utilize the signal constellations, practical relaying schemes and codes as in our  implementation, is not theoretically tractable; therefore we resort to the closest tractable theoretical models, that could still enable us to gain some intuition.

We here summarize the steps and assumptions of the theoretical
analysis; we provide more details in
Appendix~\ref{sec:app_outage}. The relevant information-theoretic
metric over Gaussian fading channels is {outage probability}: the
probability that the channel realizations do not support a target rate
$R$.  To calculate the rate $R$, we use the approximate capacity
expressions in \cite{Avestimehr2015,avestimehr_hd} for Quantize-Map-Forward (QMF), an information theoretic scheme,
that has ideologically inspired (in terms of operation) DIQIF
\cite{Quilt2014}. QMF uses Gaussian codebooks for transmission and
Gaussian quantizers to quantize each received vector at the noise
level, followed by random mapping and retransmission.  We assume
infinite complexity encoding and decoding at the network nodes. We
write expressions for outage probability using channel statistics, and
formulate an optimization problem that selects the best subset of
relays (of size $1, 2, 3$ etc.) to use over a given $N$-relay network.

We numerically solved the optimization problem over a range of
configurations, assuming Rayleigh fading; we provide indicative
plots in Fig.~\ref{fig:outage_prog_overall}, over a network with $10$
relays, where either none, or $k$ $(= 1, 2, 3, 4)$, or all relays are
chosen. Fig.~\ref{fig:outage_progression__bestworst_clus} and
\ref{fig:outage_progression_clus} consider a {clustered} topology
where some of the relays are in close range of the S--D
pair while the \iffalse{\bcomment \st{remaining relays}}\fi rest are further spread
out. Fig.~\ref{fig:outage_progression_iid} corresponds to a symmetric
setting, where all network channels have the same fading statistics.

To support {\em P1}, Fig.~\ref{fig:outage_progression__bestworst_clus} shows that the relays we select can make a significant difference, by comparing the performance of the optimal $1$ and $2$ relays (i.e., relays that leads to the best performance) to other suboptimal choices.
To support  {\em P2} and {\em P3},
Fig.~\ref{fig:outage_progression_clus} and \ref{fig:outage_progression_iid} show, over a clustered and  a symmetric configuration, the fact that the returns from increasing the number of relays diminish fast. For instance, in Fig.~\ref{fig:outage_progression_clus}, at an FER of $10^{-2}$, we see that going from $0$ relays to $2$ provides a $13$ dB gain in performance, whereas the benefit in going from $2$ to $10$ relays is $4$ dB in total. %In summary, the results in Fig.~\ref{fig:outage_progression__bestworst_clus} support {\em P1}; the results in Fig.~\ref{fig:outage_progression_clus} and  \ref{fig:outage_progression_iid} support {\em P2} and {\em P3}.

\begin{figure*}
  \centering
\subfigure[$\sigma^2_{SR_1} = 0.8\mathsf{SNR}$, $\sigma^2_{R_1D} = 0.8\mathsf{SNR}$, $\sigma^2_{SR_2} = \mathsf{SNR}$, $\sigma^2_{R_2D} = \mathsf{SNR}$, $\sigma^2_{SD} = 0.1\mathsf{SNR}$.]
  {\includegraphics[width=0.3\textwidth]{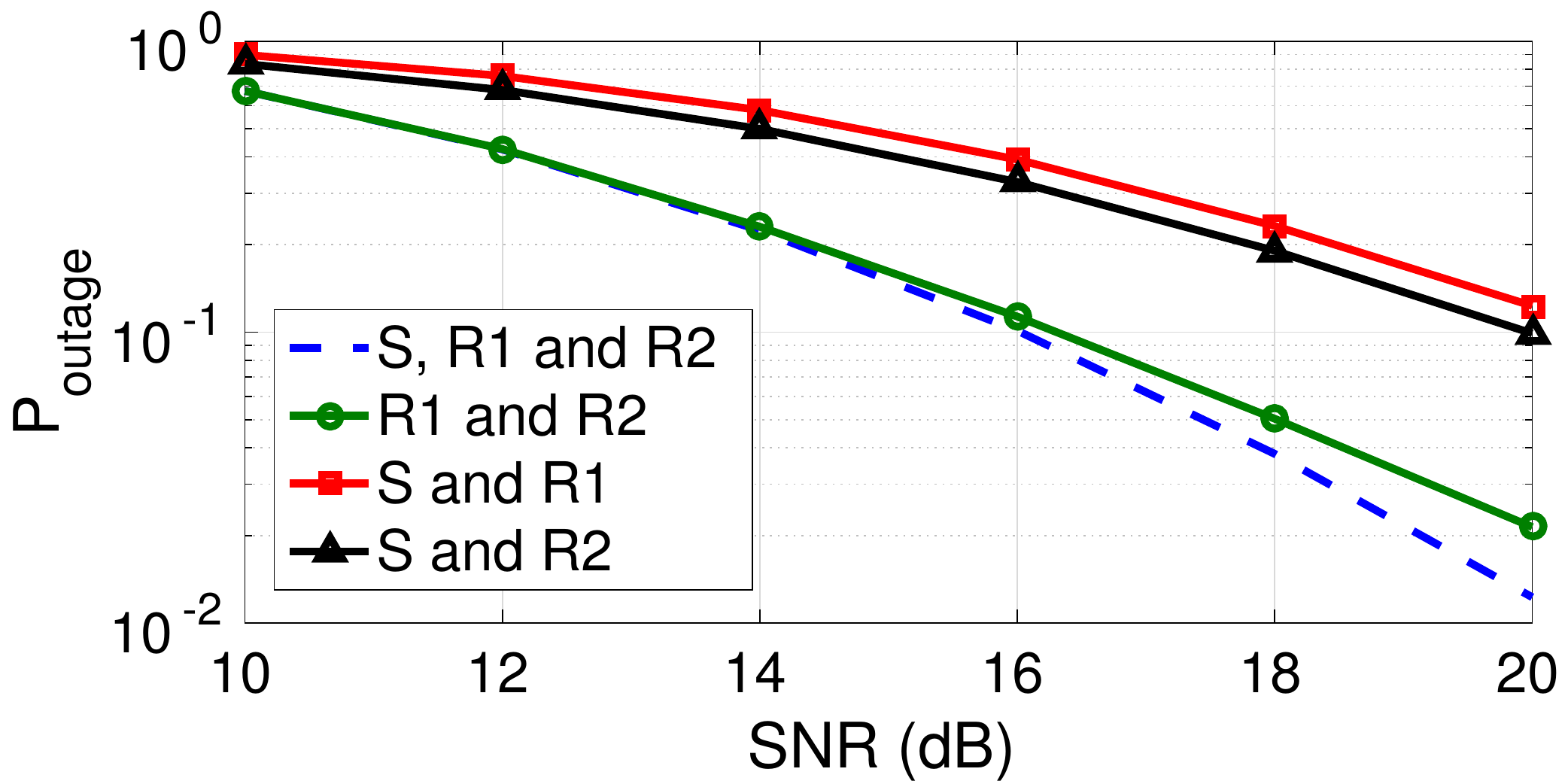}
\label{fig:outage_directlink_top1}
}
\subfigure[$\sigma^2_{SR_1} = 0.1\mathsf{SNR}$, $\sigma^2_{R_1D} = 0.1\mathsf{SNR}$, $\sigma^2_{SR_2} = \mathsf{SNR}$, $\sigma^2_{R_2D} = \mathsf{SNR}$, $\sigma^2_{SD} = 0.5\mathsf{SNR}$.]
  {\includegraphics[width=0.3\textwidth]{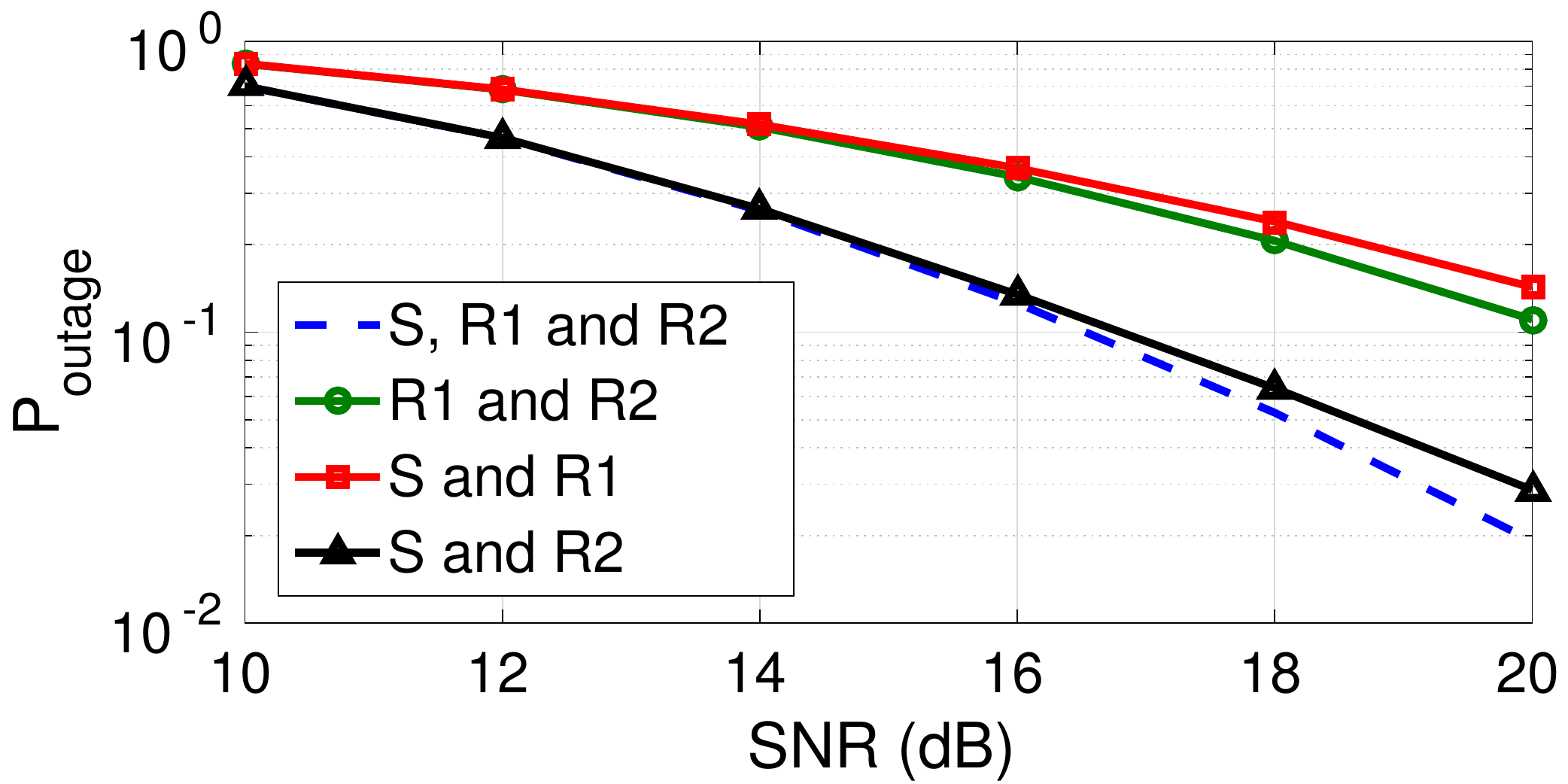}
\label{fig:outage_directlink_top2}
}
\subfigure[$\sigma^2_{SR_1} = \mathsf{SNR}$, $\sigma^2_{R_1D} = 0.9\mathsf{SNR}$, $\sigma^2_{SR_2} = \sigma^2_{R_2D} = 0.25\mathsf{SNR}$, $\sigma^2_{SD} = 0.8\mathsf{SNR}$.]{
  \includegraphics[width=0.3\textwidth]{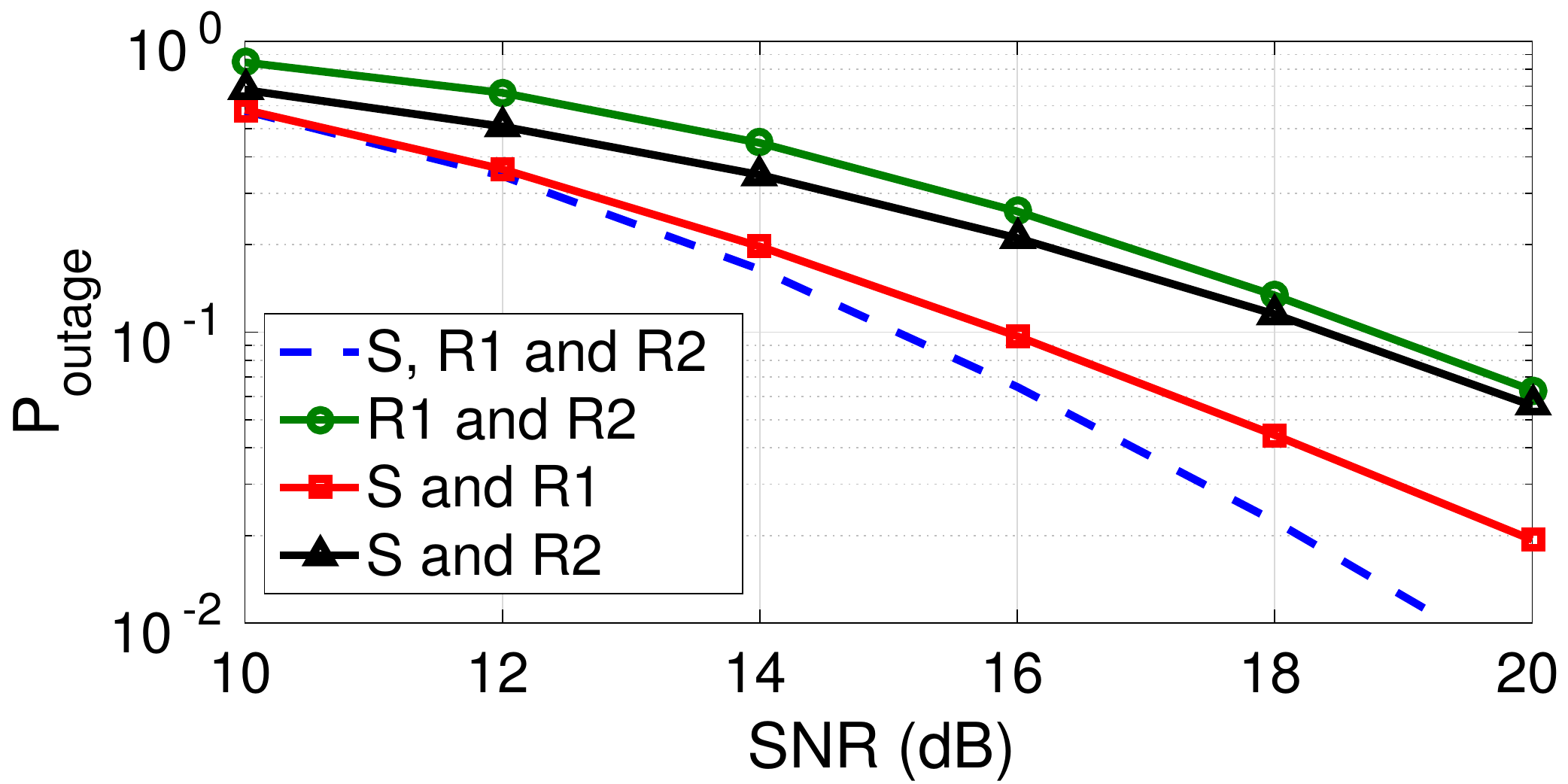}
\label{fig:outage_directlink_top3}
}
\caption{Performance of $2$-transmitter modes in Rayleigh-faded $2$-relay networks.}
\label{fig:outage_directlink}
\vspace{-1em}
\end{figure*}

To support {\em P4}, Fig.~\ref{fig:outage_directlink} compares the performance when all three nodes (S, $R_1$ and $R_2$) in a $2$-relay network simultaneously transmit, vs when any two of the nodes do so. We see that, while a particular $2$-transmitter mode can perform worse than using all $3$-transmitters, the best incurs minimal performance loss for SNR ranges of interest. Intuitively, if e.g., the S--D link is strong, then having S and the ``strongest'' relay transmit can perform very close to all three nodes transmitting; similarly, if the S--D link is weak, the $2$-relay mode approaches the optimum performance. Hence, if we can select the {best} $2$-transmitter mode, we can achieve comparable performance to that with all $3$ transmitters active.

%\vspace{-0.6em}
\subsection{Discussion} \label{sec2.4}
\paragraph*{Why DIQIF} We selected DIQIF for the relay operation for two reasons.
First, it was shown to be the best PHY strategy in 1-relay networks
\cite{Quilt2014}; in this paper, we extend DIQIF to support $2$-relay modes and
experimentally verify (in Section 5) similar trends.
Second, with DIQIF, each relay performs the same operations independently of whether it operates in
a $1$-relay or a $2$-relay mode, and independently of channel statistics.
%The fact that we maintain the same relay operations independent of the network configuration
This is an attractive attribute when we need to adapt the relays as the network changes.
We note that, although we selected a specific relaying scheme that leads to good performance, the adaptation algorithm we describe next in Section~\ref{sec:adaptive_algorithms}  could be applied over any relaying scheme, such as Amplify-Forward, Decode-Forward, etc.
% We note that although in this paper we only consider 1-relay and 2-relay modes for cooperation, our designs for the DIQIF strategy can directly be extended to  $k$-relay modes, with $k>2$. We do not do so in this paper for the reasons discussed next.

\begin{figure}
  \centering
  \includegraphics[width=0.75\columnwidth]{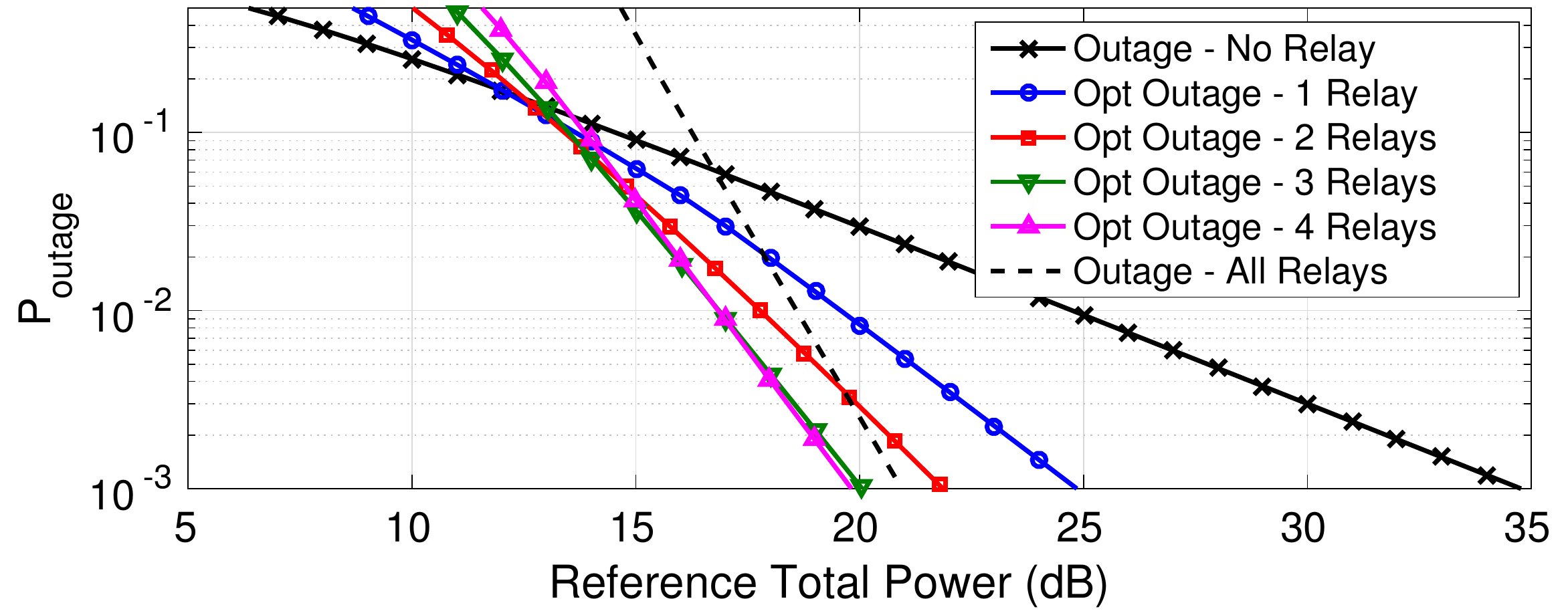}
  \caption{Power efficiency in a $10$ relay network.}\label{fig:outage_progression_totpow_clus}
\end{figure}

\paragraph*{Using more than 2 transmitter cooperation}
As Fig.~\ref{fig:outage_prog_overall} shows, the outage performance could be improved by adding more than two relays. %we here explain why using more than 2-transmitter cooperation can incur more overhead in comparison to theoretical gains:\\
However,  we believe that the overhead for more than 2-transmitter cooperation  would outweigh the benefits, as we argue next.\\
$\bullet$ \emph{Complexity of Decoding:}
Unlike setups such as MIMO, where we can perform beamforming and successively decode multiple superimposed signals, we have a single receiver antenna and a superposition of different signals that need to be simultaneously processed %(as per the QMF paradigm in \cite{ADT11})
to extract the source message. As a result, the complexity of the decoder increases exponentially with the number of superimposed signals; although we used a customized belief propagation decoder that offers the state of the art in low-complexity decoding,  more than two superposed signals led to prohibitive processing in our decoding.\\
$\bullet$ \emph{Overhead:} For PHY cooperation transmission, preamble information (training sequences and SNR estimates) needs to be transmitted from each of the $k$-transmitters to the receiver orthogonally over time. Thus, for a fixed payload size, the preamble overhead to payload scales linearly with the number of transmitters used\footnote{In Section \ref{sec:testbed}, we see that the overhead ratio in our setup is approximately $0.07k$ for $k$-transmitters.}.\\
$\bullet$ \emph{Power Efficiency:}
Having a small number of transmitting nodes is also attractive from the standpoints of interference, total power efficiency, and even total indoor emissions (health regulations).
{Fig.~\ref{fig:outage_progression_totpow_clus} plots the performance of best $k$-relay subnetworks, where the $x$-axis is normalized to the {total} power used for transmissions by all nodes in the network. We see that a total power budget is often better utilized if it is distributed among a smaller subset of well-chosen relays, as opposed to distributing it across all relays.\\}
%We see that in many cases, near-optimal power efficiency is obtained by using very few ($1$ or $2$) relays.} \iffalse{\color{green} Yahya: I think this last sentence and the plot are confusing.}\textcolor{red}{Christina: I agree it is confusing - the x-axis, can we plot-explain the figure better?}\fi\\
$\bullet$ \emph{Diminishing Returns:} Fig \ref{fig:outage_prog_overall} (without factoring in overheads) shows that the progressive benefits in going from $2$ to $3$ relays are {50 $\%$} of that in going from $1$ to $2$ relays, and {$17 \%$} of that in going from $0$ to $2$ relays.

%{ In summary, we opt not to go to 3-transmitter cooperation majorly motivated by the exponential increase in decoder complexity and the linear increase in preamle overhead as well as power-efficiency considerations.}

\iffalse
\paragraph*{More efficient implementation} \textcolor{blue}{To perhaps be moved inside the implementation section}There are several optimization that are possible to make the system more efficient. For instance, to minimize the amount of feedback we need, the source may continuously transmit several packets, i.e., aggregating multiple Phases I; the destination can then send a group acknowledgment and the relay(s) can perform the required retransmissions in an aggregated Phase II. In this paper we focused on providing a proof of concept, but we acknowledge that definitely such optimizations would be needed to have a fully fledged efficient implementation.
\fi

\section{ SPA: selecting relays}
\label{sec:adaptive_algorithms}

\noindent Given a network with $N$ relays, SPA selects which mode to activate, i.e., which one or two relays will best serve to assist the S--D communication. We assume that S transmits at a fixed rate, e.g. to support video streaming.
During selection, we use a fixed MCS (determined by the source, and also used by the relay), and employ end-to-end FER as our performance metric. Since the discovered ordering (which mode is better) extends to any fixed MCS, with rate adaptation (on top of mode selection), the best mode still achieves a better FER.
\subsection{Main idea}
% We do so by making use of destination FER-statistics to ``learn'' a mode in the network among the possible cooperation modes through adaptive training and feedback, using a machine learning approach.
SPA starts by identifying the mode that achieves the lowest FER among all $N$+${N}\choose{2}$ modes. Subsequently, SPA is triggered whenever the experienced FER of the selected mode exceeds a predefined threshold. The main design challenge is efficiency: even for a moderate $N$, learning across all modes at the destination may incur significant overhead - both in terms of rate and in terms of the ``mode switches'' required to test the modes and converge. We make SPA efficient in three ways:\\
$\bullet$ We use information-carrying frames for training, i.e., we send information messages via each of the modes and use the FER of the received information messages\footnote{We estimate the FER by decoding and checking the CRC.} to decide on the winner mode during learning. We refer to these as training frames in the following, yet we emphasize that there is {\em no rate overhead}; only the performance penalty of not always utilizing the best mode.\\
$\bullet$ Instead of sending the same number of training frames for
each mode, we progressively weed out modes with
high FER and thus  allocate more training resources to
the modes that are likely to be selected. Thus we converge faster to
the mode with the lowest FER.\\
$\bullet$ SPA keeps memory--in the form of ranking of modes after each learning phase. The intuition is that if the configuration changes, the ``now best'' mode is likely to be found among the previously stronger modes.
%unlikely to be a mode that was  extremely weak; it is more probable to be found among the previously stronger modes.
Accordingly, every time we need to adapt, we use the ranking of the modes in the previous learning phase to train over a smaller \emph{partition} (subset) of modes. This reduces the number of ``mode switches'' required, and as we show in Section~\ref{sec:experimental_results} (Fig.~\ref{fig:algo_memory}), suffers from negligible FER-penalties. Also, it is this facet of the algorithm that gives it its name: \textbf{S}ort, \textbf{P}artition and \textbf{A}dapt.

\subsection{Algorithm description}
%============================================
SPA is described in pseudo-code in the following page. For clarity, we separately describe the function LEARN that searches for the best mode starting from a set of candidates; SPA invokes LEARN to adapt as needed, and provides to LEARN the starting set of mode candidates.
%The algorithm we use to `learn', or converge to the best mode, based on destination FER, is illustrated in the form of a pseudo-code, broken for clarity in  two parts, in  Algorithms \ref{algo:code} and 2. Algorithm 1 describes the search for the best mode starting from a set of candidates; Algorithm 2 evokes Algorithm 1 to adapt as needed, and provides to Algorithm 1 the starting set of mode candidates.
%\textcolor{blue}{Does the current pseudocode include the memory (can we include it)? Can we explicitly show what the input  parameters of this algorithm ? can you please also add a table that describes what are the variables in the pseudocode.}
SPA is inspired from a popular machine learning approach of \emph{learning from the best expert} \cite{bestexpert}, appropriately customized for our setup. Two main differences are that here we are looking to \emph{converge} to the best expert's FER, where an expert corresponds to a mode; and that we use memory to adapt to network changes.

\renewcommand{\thealgorithm}{}
{
\begin{algorithm}
\small
\floatname{algorithm}{LEARN}
\begin{algorithmic}
\renewcommand{\algorithmicrequire}{\textbf{Input:}}
  \REQUIRE Set of $r$ modes $\{m_1,\cdots,m_r\}$.
\renewcommand{\algorithmicrequire}{\textbf{Output:}}
  \REQUIRE Ranked list of $r$ modes.
 % \STATE Initialize batch size $l$, learning rate
 % $\eta$, shift parameter $\alpha$, threshold $\epsilon$, max
 % batches $B$.
  \STATE $C_0 \leftarrow \{m_1,\cdots,m_r\}$, $\forall i \in C_0, w_i
  \leftarrow 1/r$, $j\leftarrow 0$, $L'\leftarrow \{\}$.
  \WHILE{($|C_j|>1$ and $j<B$)}
  \STATE Run the modes in $C_j$ for $l$
  frames.
  \STATE Compute the empirical FER for mode $i$ as
  $\hat{f}_{ij}$.
  %, for all modes in $C_j$.
  \STATE $P \leftarrow 0$,
  $tot \leftarrow 0$, $C_{j+1}\leftarrow \emptyset$ \FOR {$i \in C_j$}
  \STATE $w_i \leftarrow w_i e^{-\eta \hat{f}_{ij}}$
  \STATE $P \leftarrow P + (1-(1-\alpha)^{\hat{f}_{ij}})w_i$
\ENDFOR
\FOR {$i \in C_j$}
\STATE $w_i \leftarrow (1-\alpha)^{\hat{f}_{ij}}w_i + \frac{1}{n-1} (P- (1-(1-\alpha)^{\hat{f}_{ij}})w_i)$
\STATE $tot \leftarrow tot + w_i$
\ENDFOR
\FOR {$i \in C_j$ in ascending order of $w_i$}
\STATE $w_i \leftarrow w_i/tot$
\IF {$w_i > \epsilon$}
\STATE $C_{j+1} \leftarrow C_{j+1} \cup \{i\}$
\ELSE
\STATE $L' \leftarrow \text{append}(L',\{i\})$
\ENDIF
\STATE $j\leftarrow j+1$
\ENDFOR
\ENDWHILE
\STATE Sort $C_j$ in ascending order of weights.
\STATE Append $C_j$ to $L'$ and output reverse($L'$).
\end{algorithmic}
\caption{Find the best mode}
\label{algo:code}
\end{algorithm}
}
\noindent{\bf Learning in batches and weight update.}
LEARN starts from a set of admissible modes.
%; initially all modes  are admissible.
There is a weight $w_i$ for each mode $i$.
%  has associated with it
%a weight $w_i$; the weights are used to rank the modes.
Initially all weights are equal, with their sum normalized to 1.
Learning takes place in \emph{batches} of frames. In each batch, $l$ frames per
admissible mode are sent for training (we emphasize  that these frames
encode and convey information messages to the destination; they are not
dumb training frames).
After each batch, the weights of the admissible modes are updated.
A mode with high FER is penalized exponentially in terms of its weight;
the {learning rate} $\eta$ decides the rate of penalization.
The shifting parameter $\alpha$  enables robustness against sudden
changes in FER.
% during the learning phase.

\noindent{\bf Early Reject of Bad Modes.}  After each batch is processed, the
modes with $w_i$'s less than a threshold $\epsilon$ are rejected.
%shrinking the admissible set from one batch to the next.
This ensures that (i) bad modes are {not} retained throughout the learning phase, and (ii) when
a clear winner exists, faster convergence to the winner.

\noindent{\bf Mode Selection and Termination.} Learning can at most continue for $B$ batches,
after which a hard decision is taken by selecting the mode with the highest weight.
The algorithm may converge before $B$ batches if  \emph{exactly one} mode remains in the learning phase. {We set the value of $B$ online during training, to a value for which the algorithm converges; this incurs no extra overhead.}
% \textcolor{green}{A reviewer asked how do we select B? We should write something.}\textcolor{red}{A: We did this choice by taking streams from the data, and observing the FER obtained \emph{after} the training phase. With very low batchsize, it converges to wrong modes, and hence the post-training FER is very high. The plot of post-training-FER versus number of batches sharply falls upto a point, after which it tends to saturate, and even goes up slightly, since we continue to train over modes other than the best. The process for parameter selection in most such machine learning algorithms (to my limited knowledge, and upon past discussions with SB while doing this) is eyeballing/heuristic based, as opposed to being from set formulae.}
\iffalse
While progressively weeding out the bad
modes as described above, if we are left with only one mode in the
admissible set after $b$ batches $(b < B)$, then we declare that mode
as the winner.  If however, after $B$ batches, there is more than one
mode in the admissible set, we declare the mode with the highest
weight as the winner.
\fi
 %and allow that mode to operate before the next algorithm
%refresh is due.

\noindent{\bf Adapting to configuration changes.} LEARN is used by SPA to adaptively select the best mode of operation based on a triggering mechanism. LEARN is invoked whenever the ``windowed FER'' (i.e. FER in the last $w$ frames) of the currently used mode exceeds a pre-determined threshold ($\zeta$).
%,  to find a mode that  meets the FER requirements.
%This enables to adapt to configuration changes that often take place with wireless
%transceivers.

\noindent{\bf Learning over ranked subsets of modes.}
Whenever we invoke LEARN, we use the ranking of modes in the previous cycle to select a subset of ``high-ranked'' modes over which we run our algorithm. The subset size ($r$) is an algorithm parameter, providing a tradeoff between accuracy (better for larger $r$) and switching overhead (better for smaller $r$). We maintain a ranked list $L$ of modes, approximately sorted in ascending order of FER. Whenever the FER of the last $w$ frames, which itself is computed at intervals of $\Delta w$ frames, drops below $\zeta$, we do either of two things: if the new trigger happened very quickly after the previous one (this is controlled by a parameter $s$), the top $r$ modes in $L$ are pushed to the end of the list; otherwise we only push the top mode to the end of $L$. In both cases, LEARN is called with the $r$ modes on top of $L$, and their ranking is updated using the output of LEARN. After every invocation of LEARN, the top mode in $L$ is used to operate the network. Table~\ref{tab:algo_terms} collects the parameters used.

{
\renewcommand{\algorithmicrequire}{\textbf{Input:}}
\begin{algorithm}
\small
\floatname{algorithm}{SPA}
\begin{algorithmic}
%  \REQUIRE Set of modes $M$ and FER threshold $\zeta$.
%  \STATE Initialize fer window size $w$, memory size $r$, minimum number of windows $s$.
  \STATE Maintain $L$ as list of modes in $M$.
  \WHILE{End of Time}
  \STATE $i\leftarrow 0$.
  \STATE Run network using mode $L[1]$ for $w$ frames.
  \WHILE{True}
    \IF{Empirical FER in last $w$ frames $\geq \zeta$}
        \STATE Break.
    \ENDIF
    \STATE Run network for $\Delta w$ frames using mode $L[1]$.
    \STATE $i \leftarrow i+1$.
  \ENDWHILE
  \IF{$i\leq s$}
      \STATE $L \leftarrow \text{append}(L[r+1\hdots |M|],L[1\hdots r])$.
  \ELSE
      \STATE $L \leftarrow \text{append}(L[2\hdots |M|],L[1])$.
  \ENDIF
  \STATE Call \textbf{LEARN} with $L[1\hdots r]$ and update top $r$ entries of $L$ with its output.
\ENDWHILE
\end{algorithmic}
\caption{Adaptive learning}
\label{algo:code_outer}
\end{algorithm}
}
\begin{table}[th]
    \small
    \centering
\begin{tabular}{|l|l|}
\hline
%\multicolumn{2}{|l|}{SPA}\\ \hline
$M$ & {\small Set of modes where $|M| = N + {N\choose 2}$.}\\ \hline
$\zeta$ & FER threshold \\ \hline
$r$ & Memory size \\ \hline
$w$ & Size of window to calculate empirical FER \\ \hline
$\Delta w$ & Number of additional frames \\ \hline
$s$ & Minimum number of windows\\ \hline
%\multicolumn{2}{|l|}{LEARN}\\ \hline
$l$ & Learning batch size\\ \hline
$\eta$ & Learning rate\\ \hline
$\alpha$ & Learning shift parameter\\ \hline
$\epsilon$ & Learning threshold\\ \hline
$B$ & Maximum number of learning batches\\ \hline
\end{tabular}
\caption{Parameters used in \textbf{SPA} and \textbf{LEARN}}
\label{tab:algo_terms}
\vspace{-1em}
\end{table}
\subsection{Discussion}
\label{subsec:FER_vs_cutset}
%%SPA experimentally learns the FER of different modes to select the best.
%\noindent{\bf Why not predict from channels?} Why not  learn the channel statistics in each of the modes, and then select the mode that has the
%best outage performance? Indeed, this approach would work
%well if we were comparing point-to-point wireless channels, where we have exact capacity and outage characterizations (directly computable from the channel statistics) as well as practical coding schemes that achieve the capacity.  Unfortunately this is not the case even for small (one or two relay) networks that employ
%physical layer cooperation. Information theory only gives us {bounds}, or at best some {approximate expressions} on the capacity/outage; we can use these to get theoretical insights, yet we do not have practical schemes that achieve them. %Real world effects like channel estimation errors also add to the problems with channel-based predictions in networks.
% Indeed, from our experiments, we found that if we compare modes using channel statistics and mathematical outage expressions we cannot accurately predict which are the best modes in terms of the actual FER performance observed. Fig.~\ref{fig:channel_fer} plots the observed  versus channel-predicted FER
% ``theoretical expected rates'': if the channels enabled a good predictor, we would have seen a {monotonically decreasing} scatter plot pattern, which is not what we observe.
{\noindent{\bf Why use FER-driven learning?}
We consider part of our contributions, that using FER-driven learning in SPA is an educated choice: we put effort in exploring and ruling out alternative methods. We can think of three basic approaches for selecting relays: distance-based, channel-based and FER-based. For PHY cooperation, distance-based selection turns out to be too simplistic: it does not give consistent performance over wireless, due to effects such as interference, obstructions, multipath fading, etc. The second alternative is to learn the channels between all the network nodes, and use that knowledge to select relays, for instance, by evaluating theoretical rate expressions over the subnetworks. This approach, that works well for packet routing over relays, is surprisingly not reliable when selecting subnetworks for PHY cooperation. We ruled out this alternative because, from the channel knowledge, we can only calculate theoretically {\em approximate} expressions of the relay network performance; we found that the ordering (which subnetwork is better) indicated by these approximate expressions is not consistent with the ordering achieved by the actual scheme in practice.}

\begin{figure}
\centering
\subfigure[Frame errors over time.]
{\includegraphics[width=0.47\columnwidth]{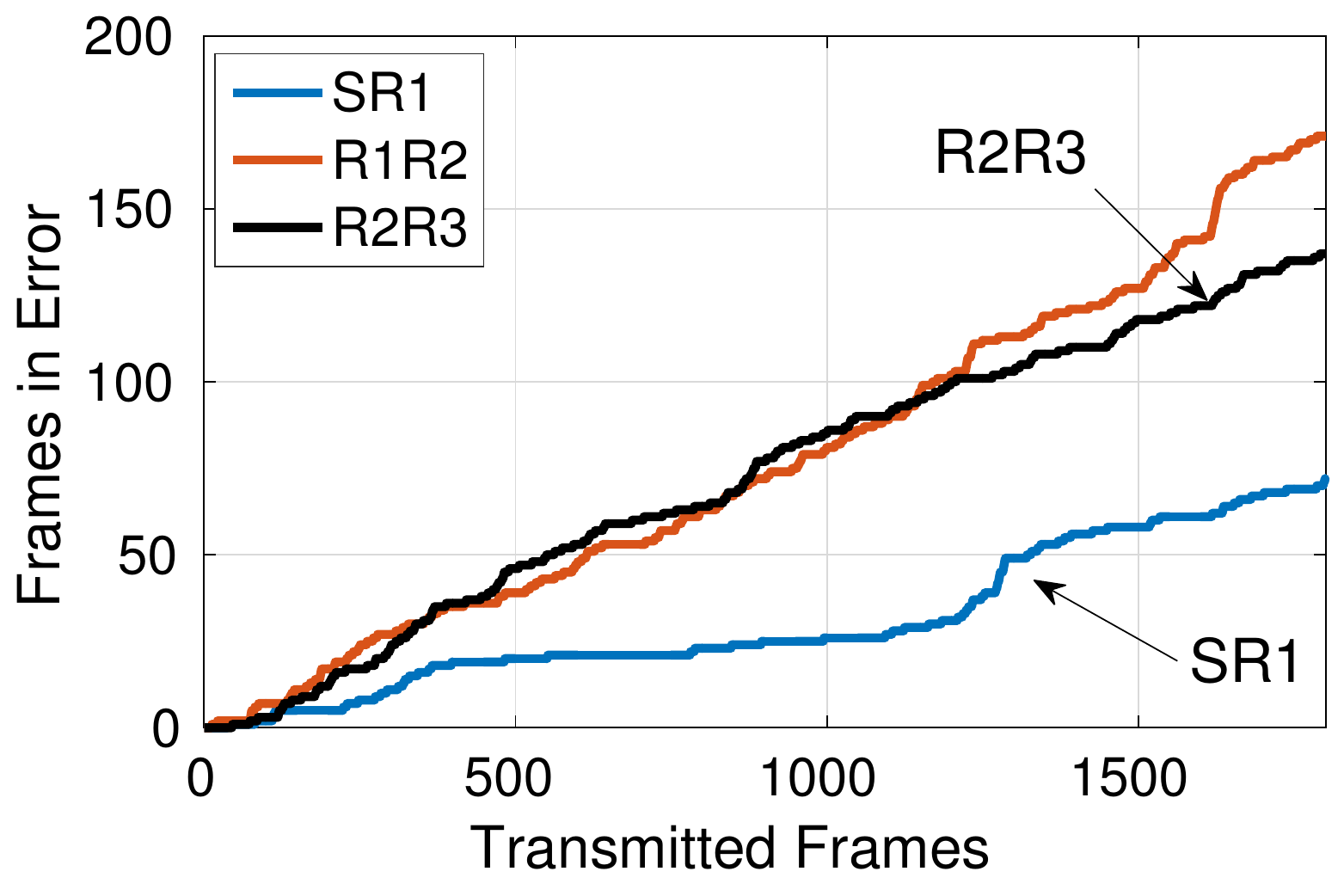}
\label{fig:frameerrors_ch_vs_fer}
}
\subfigure[Channel-predicted rates.]
{\includegraphics[width=0.47\columnwidth]{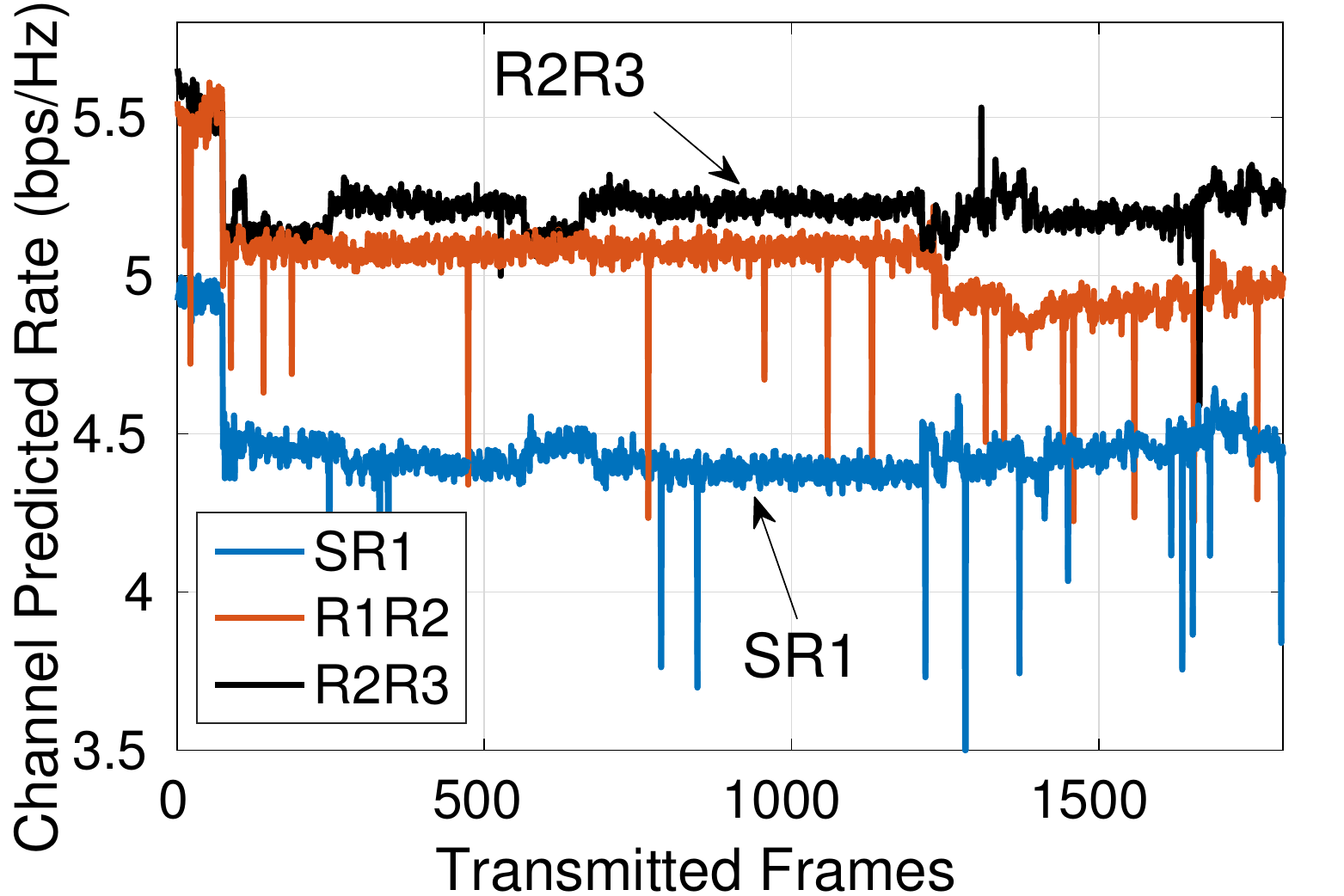}
\label{fig:channelpredictions_ch_vs_fer}
}
\caption{FERs vs channel-based predictions.}
\label{fig:channel_fer}
\vspace{-1em}
\end{figure}

{For example, in Fig. \ref{fig:channel_fer}, we see that over this static configuration, mode $R2R3$ is theoretically predicted (using channel estimates from the implementation) to be much better than $SR1$ and could support a $0.5$ bps/Hz higher rate. However, looking at FER, we see  that in fact it for $R2R3$, it is approximately $2$x that of mode $SR1$.
The inconsistency in the example is because, in contrast to point-to-point links, there does not exist exact information-theoretic capacity expressions for PHY cooperation networks; only bounds on capacity are available, and the gaps are in general channel and topology ($1$ vs $2$ relay(s)) dependent. Explicit theoretical characterizations of the QMF rate that take into account the quantization, mapping and channel estimation errors that occur in implementations such as ours, have so far remained elusive.
Finally, this approach (similarly used in \cite{Jing2009,Shi2008,LoHeath2009}) comes with the overhead of collecting (and sharing) accurate channel information of all channels, which is especially high in OFDM-based systems like WiFi, where we need channel information per subcarrier, per link.}

\noindent{\bf How does SPA scale?}
%A core idea in {SPA} is that, even if we have a very large network, we can achieve consistent performance at low overhead
%by focusing our training  where it matters, over a small number of modes -- ones learned from the past, or identified
%through training if the current set fails. Thus,
 {SPA} can be applied over arbitrarily large networks, since its complexity
is related to the number of modes invoked in LEARN, as opposed to the total number of modes in the network, which is $N + {N \choose 2}$ for an $N$-relay network.\footnote{Having said that, we do not expect in an indoor environment (our application focus) an arbitrarily large number of  available relays within range.} We found in our experiments that we do not need to train over more than $3$ or $4$ different modes at a time. For instance, in Section~\ref{subsec:expt_performance_adaptive} (see Fig.~\ref{fig:algo_memory}) we show that over $3$ relay networks, a subset size of $3$ modes (as opposed to the total of $6$ modes) at each learning epoch allows to extract excellent FER performance while also minimizing the number of mode-switches necessary, achieving faster convergence and less feedback overhead.\\
{\noindent{\bf Why not brute force search?}  Even with a small number of relays, SPA can significantly outperform exhaustive search (0.5x FER), as we show in Section \ref{sec:experimental_results}.}

\section{System Implementation}
\label{sec:system_implementation}

\subsection{Physical Layer }
\begin{figure}[t]
  \centering
  \includegraphics[width=0.9\columnwidth]{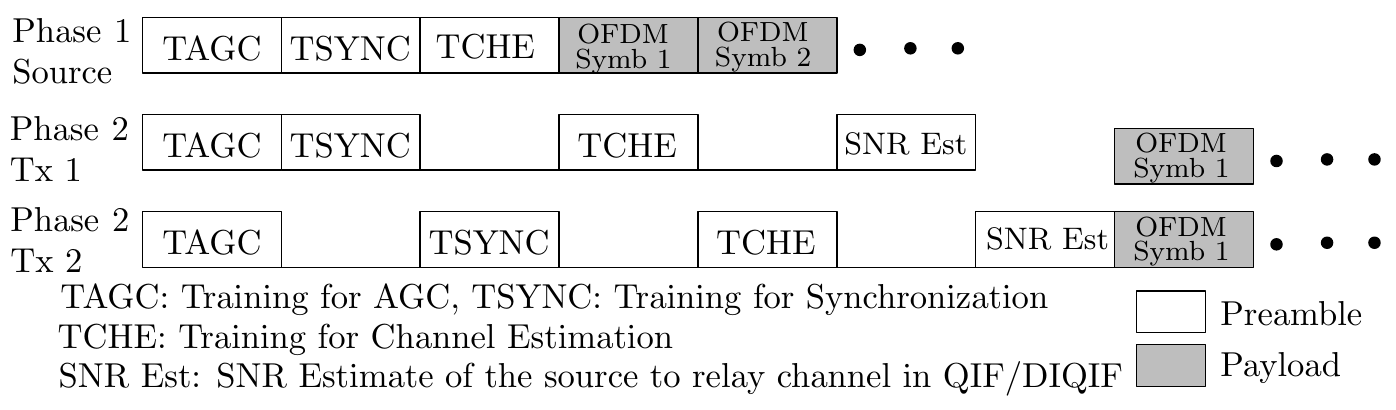}
  \caption{Time diagram.}
\label{fig:timing_diag_ipe}
\end{figure}

%-----------------------------
\noindent We follow the PHY procedures of
WiFi (IEEE 802.11).  Each transmitted frame consists of a preamble and
a payload, as illustrated in Fig. \ref{fig:timing_diag_ipe}.

\noindent \textbf{Preamble.}
The preamble consists of TAGC, TSYNC, and TCHE OFDM symbols structured as explained in Fig.~\ref{fig:timing_diag_ipe}. The TCHE OFDM symbol is additionally used for CFO estimation.
%The training sequences used in the preamble are explained in Fig.~\ref{fig:timing_diag_ipe}.
%The TCHE OFDM symbol is additionally used for CFO estimation.
In Phase 1, only the source transmits, using the structure in Fig.~\ref{fig:timing_diag_ipe}. In Phase 2, TAGC is sent by the two transmitters simultaneously, with a cyclic shift between them to avoid accidental nulling. TSYNC and TCHE are orthogonalized in time. For $k$ cooperating transmitters, the preamble therefore consists of $3k+1$ OFDM symbols.

\noindent\textbf{Payload.}
The payload consists of 42 OFDM symbols, i.e., data and pilot subcarriers as in 802.11.
In Phase 1, for all  schemes, we transmit the payload  corresponding to an OFDM-based single Tx single Rx antenna system.
In Phase 2,  the source (if selected to transmit) retransmits the same payload as in Phase 1, and the relay(s) transmit processed signals. The preamble-to-payload ratio is therefore 0.07k for $k$ cooperating transmitters.

\noindent\textbf{Synchronization.}
For Phase 2 transmissions, carrier and timing synchronization is performed through a wired connection between nodes --the same approach as presented in~\cite{MobihocQMF}. Yet, we would like to mention that work on distributed transmissions has shown the viability of achieving accurate timing and carrier synchronization in a distributed manner \cite{RiceWork,SourceSync,AirSync}; these are enabled by implementing a large part of the mechanisms in real time in the FPGA to achieve fast turnaround times. We did not incorporate this into our testbed, where we focused on proof-of-concept experimentations.

\subsection{Testbed} \label{sec:testbed}
We used the WARP SDR hardware to implement the source, relays and
destination nodes  and the WARPLab framework to interact
with the hardware via a host PC running MATLAB. The samples to be
transmitted by a node were generated in MATLAB and downloaded to the
transmit buffer of the corresponding node. The host PC triggered a
real-time over-the-air transmission and reception by the nodes. The
samples received at a node were read by the PC and processed in
MATLAB.

\begin{figure*}[t]
\centering
\subfigure[Per-link RSSIs for the topologies.]
  {\includegraphics[width=0.65\columnwidth]{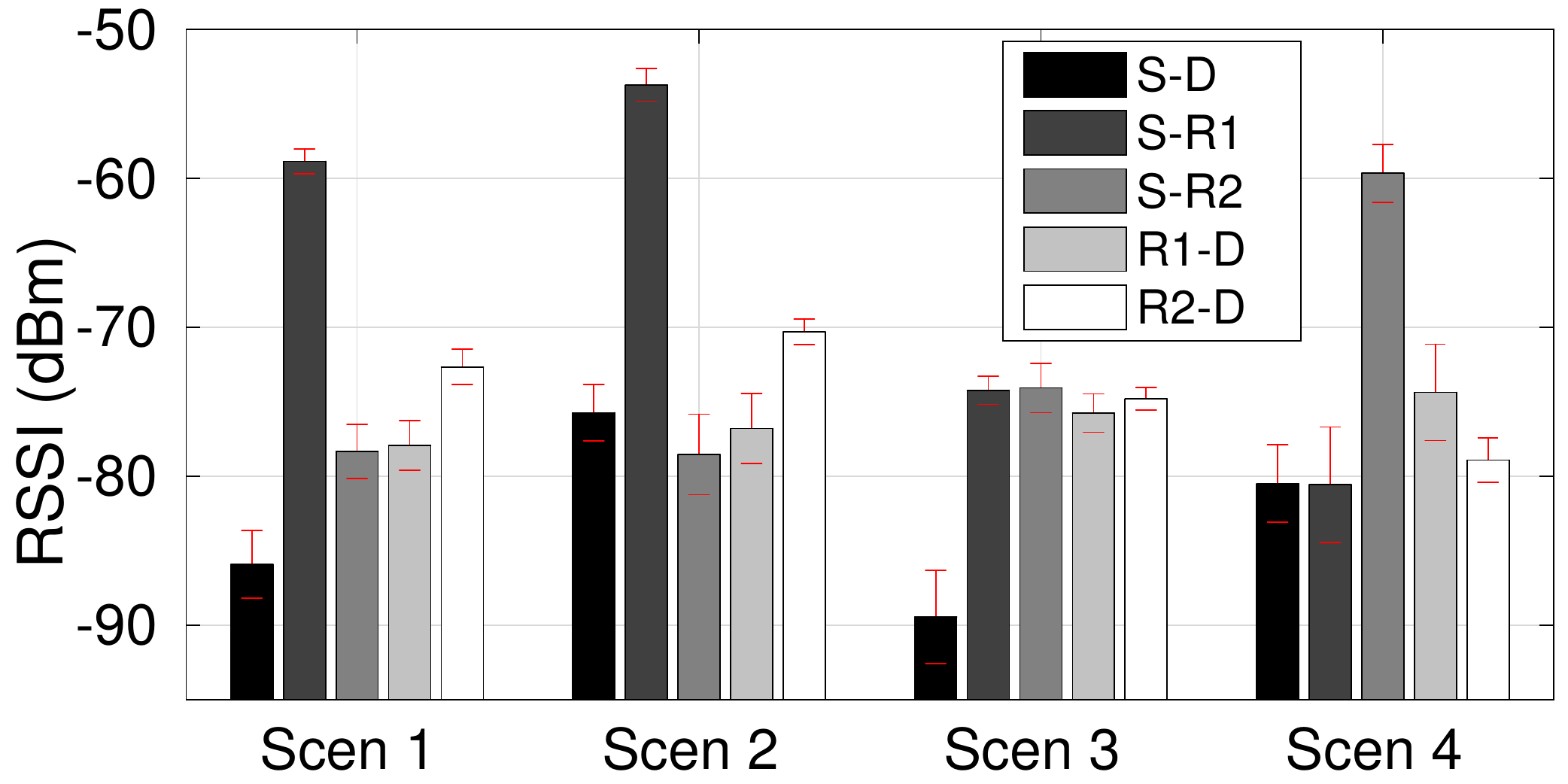}
\label{fig:rssi_per_link}
}
\subfigure[DIQIF performance in $2$-relay mode.]{
  \includegraphics[width=0.65\columnwidth]{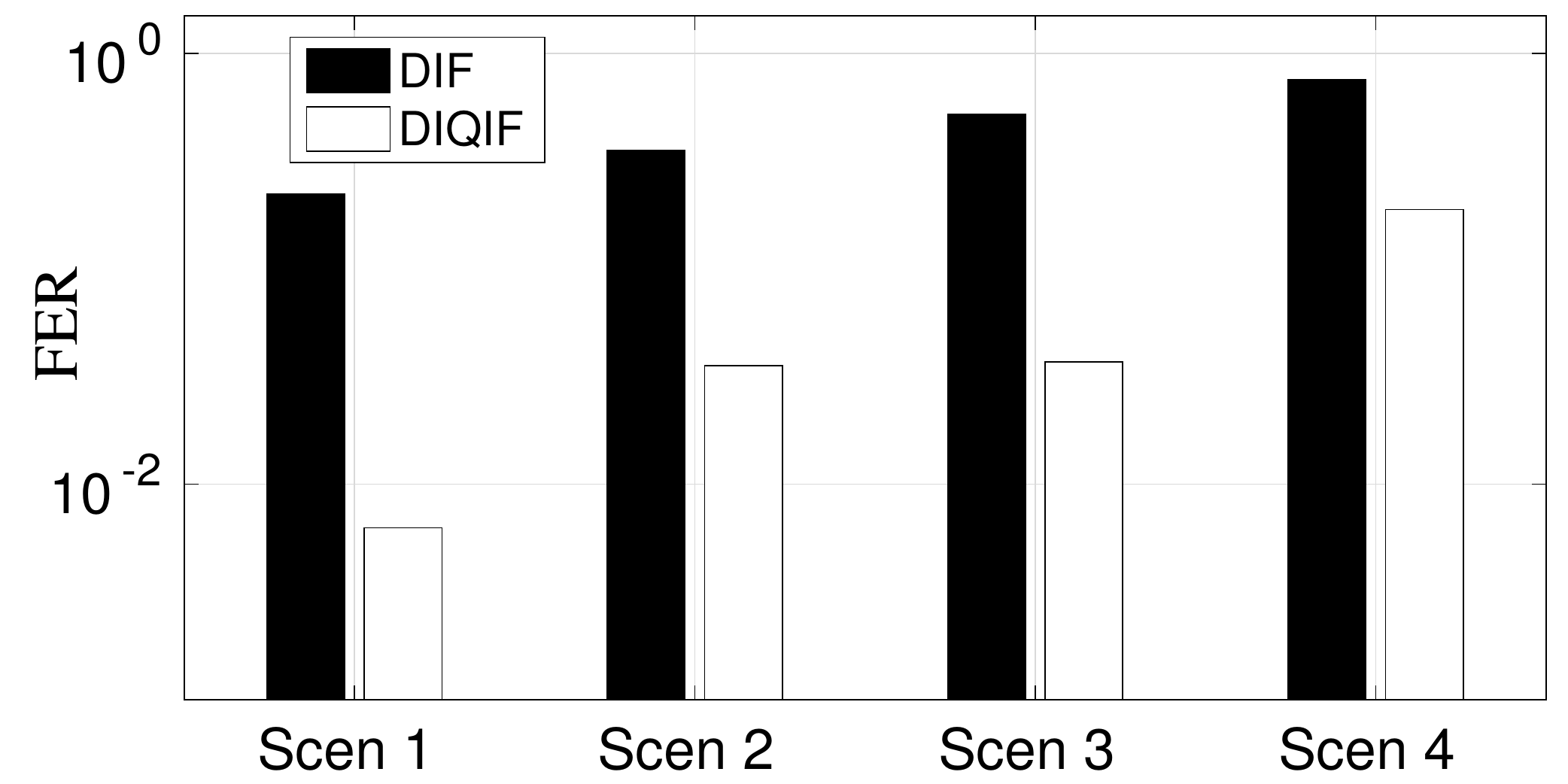}
\label{fig:FER_tworelay}
}
\subfigure[FER variability across modes.]
{\includegraphics[width=0.65\columnwidth]{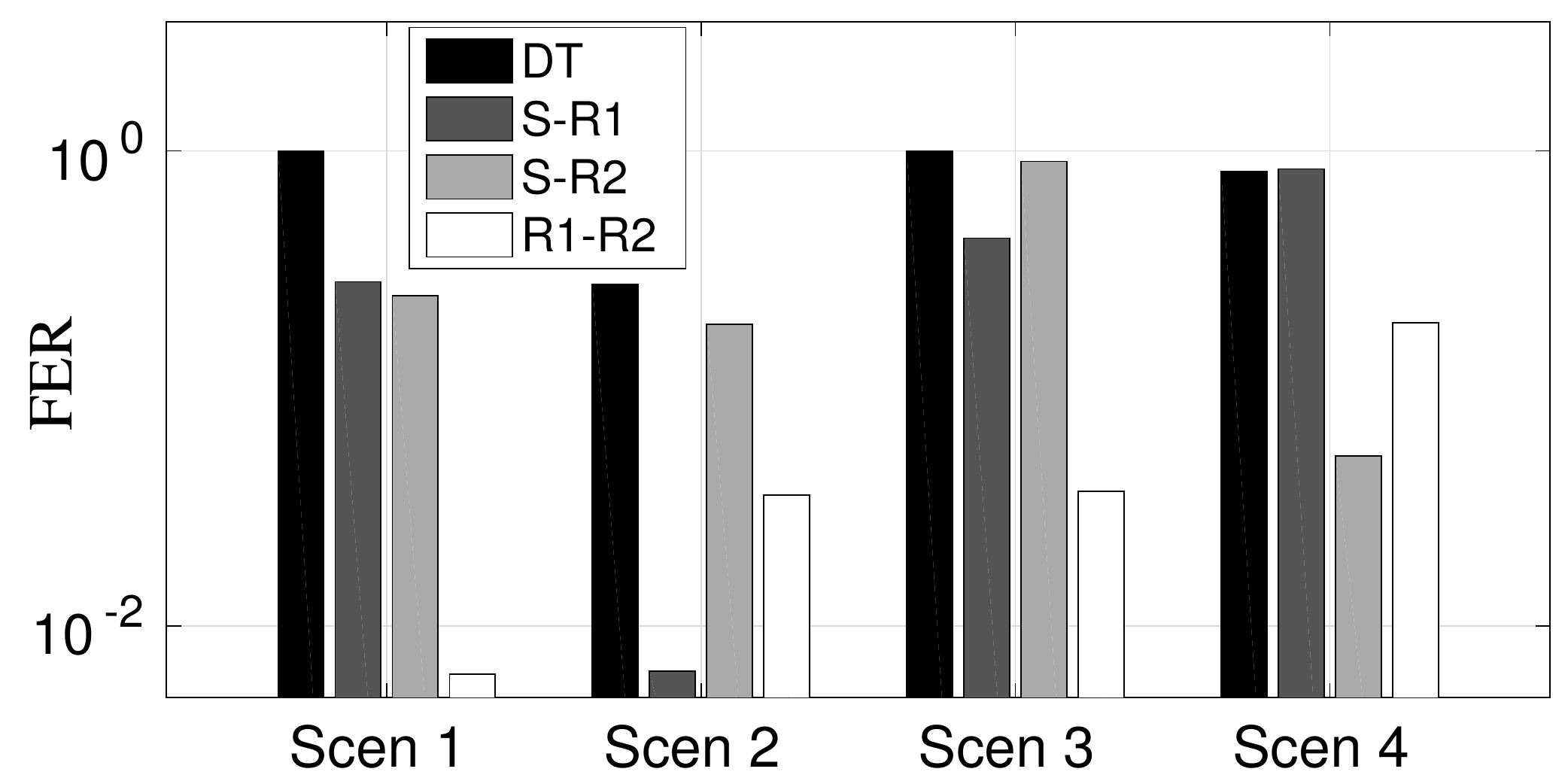}
\label{fig:FER_average_adaptive}
}
\caption{Results for $2$-relay networks.}
\label{fig:2rel_networks}
\vspace{-1em}
\end{figure*}

\section{Experimental Evaluation}
\label{sec:experimental_results}
\subsection{Performance over $2$-relay networks}
\label{subsec:results_2relay}
We created $4$ scenarios, each having a source S, a destination D and two relays R1 and R2. We report the Received Signal Strength Indicator (RSSI) values for each scenario in Fig. \ref{fig:rssi_per_link}; RSSI variations are due to distance, multipath, and transmit power adjustments. For each setting, we ran experiments for at least $1500$ coded frames. We used $16$-QAM constellations with a coding rate of $5/6$. In scenarios 1 and 2, we have R1 close to the source and R2 close to the destination; the difference is that in scenario 1 the S-D link is very weak. In scenario 3 the direct S-D link is again weak while the rest of the links are of approximately the same strength. In scenario 4 we have a strong  S-R2-D path.

We compare: (i) {\em Decode-Interleave-Forward (DIF):} Decode at the relay if possible and transmit the bit-level interleaved signal. If relay decoding is not possible, the relay remains silent.
(ii)  {\em Decode-Interleave-Quantize-Interleave-Forward (DIQIF):}  DIF if relay decoding succeeds; quantize to constellation, interleave and forward (QIF) otherwise.
(iii)  {\em Direct Transmission (DT)}, where in Phase 2, the source repeats the Phase 1 signal. %Our findings are summarized below:

\noindent{\bf (1) DIQIF performance in 2-relay modes.}
Fig.~\ref{fig:FER_tworelay} looks exclusively at the
  2-relay mode, and compares the performance of DIQIF with that of DIF, in terms of FER. In keeping with the trends from $1$-relay networks in \cite{Quilt2014}, DIQIF exhibits consistently better performance than DIF. This is because, even when the relay cannot decode, with physical layer cooperation (and DIQIF) it can still forward useful (quantized) information that can help the receiver decode; in contrast, schemes like DIF,  or simple routing along paths, require that the relay itself decodes, creating an unnecessary  bottleneck.

  %thanks to the gains obtained from forwarding usable quantized information by the relay, even when it can't decode.
  %This is a key feature that we exploit in our system, that can only be exploited when employing physical layer cooperation; treating relays as part of line networks (for e.g, in routing) often ends up making relay-decoding the performance bottleneck.\\

\noindent{\bf (2) $0$, $1$ or $2$ relays? Which ones?}
 Fig.~\ref{fig:FER_average_adaptive} shows the FERs if
we operate: (i) using only the direct S-D link, (ii) constantly in
1-relay mode with relay R1, (iii) constantly in 1-relay mode
with relay R2, and (iv) constantly in 2-relay mode with
both R1 and R2. We report these for each of the $4$ scenarios. We find:\\
$\bullet$ {\em Cooperation offers benefits over direct transmission.} As
shown in Section 3, the maximum gains are obtained
during the first steps of cooperation, i.e., including $1$ or $2$
relays, as opposed to none. Fig. \ref{fig:FER_average_adaptive} shows in some cases a
\emph{greater than two orders of magnitude benefit} in FER of the best cooperative mode over using DT.\\
$\bullet$ {\em We cannot select one mode to operate across all configurations.} This is because: (i) There is no mode that is universally better or is universally worse. In the spectrum of configurations (not all reported here), we found all modes to dominate in at least some particular scenario. (ii) The $1$-relay mode can perform better than the $2$-relay mode; this was the case in scenarios $2$ and $4$.  This could happen %when %one of the relays receives very noisy signal, or
 when the S--D link is strong.
%This happens when
%one of the relays receives a noisy, and the source (which always has a clean copy of the message) is a better transmitter in Phase $2$ that the noisy relay.
\begin{figure}[t]
\centering
  \includegraphics[width=0.95\columnwidth]{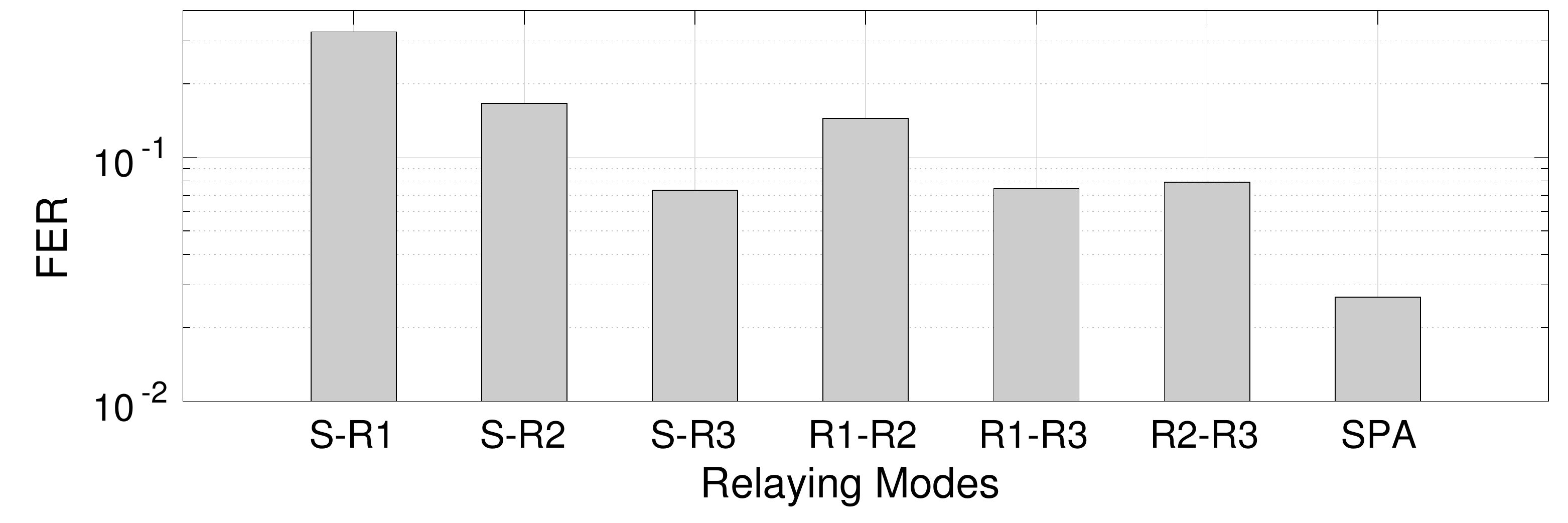}
  \vspace{-1.3em}
\caption{Adaptation in a time-varying network.}
\label{fig:adaptation_helps_overall}
\vspace{-1em}
\end{figure}
\subsection{Adaptation provides benefits}\label{subsec:results_adaptation}
Fig. \ref{fig:adaptation_helps_overall} demonstrates the benefits of adaptive mode selection in a $3$-relay network, when the topology of the network undergoes multiple changes over time. In this setup, we had $4$ physical topology changes, and each topology was held constant for $172$ frames. The relaying operation used was DIQIF, coding rate was $5/6$ and modulation was $16$-QAM. The algorithmic parameters for SPA were: $\zeta = 0.1$, $r = 3$, $w = 40$, $\Delta w = 1$, $s = 3$, $l = 1$, $\eta = 3$, $\alpha = 0.4$, $\epsilon = 0.05$, $B = 50$.
We see that
%As we can see from Fig. \ref{fig:adaptation_helps_overall},
 no mode without adaptation can on its own achieve the FER performance of SPA. This is because, which mode is best depends  on the topology of the network, which changes over time -- and SPA adapts to these changes.
 %by learning the mode best suited to the current topology.

\begin{figure*}
\centering
\subfigure[FER performance of variants.]{
  \includegraphics[width=0.65\columnwidth]{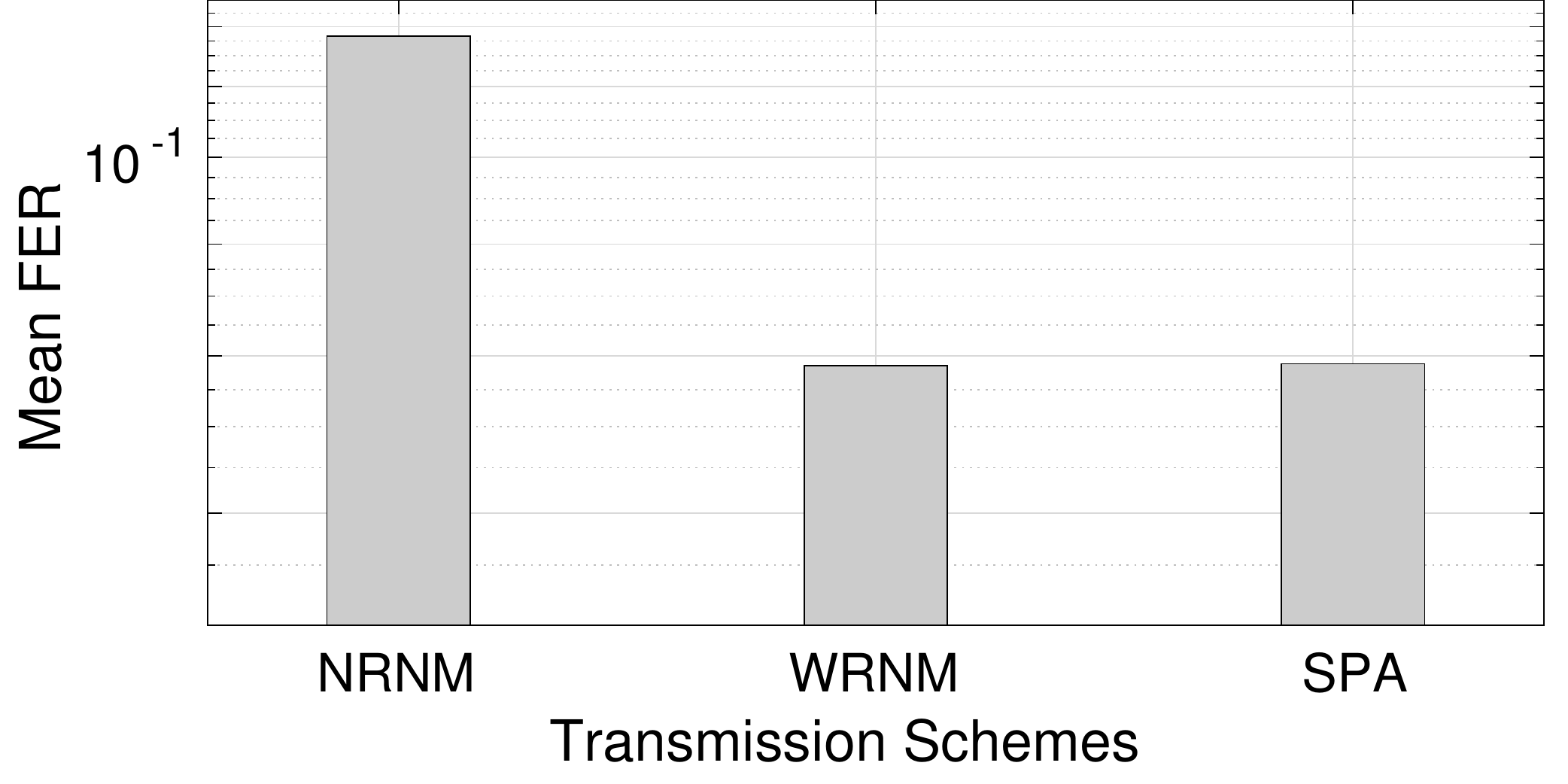}
\label{fig:algovariants_fer}
}
\subfigure[Switching performance of variants.]
{\includegraphics[width=0.67\columnwidth]{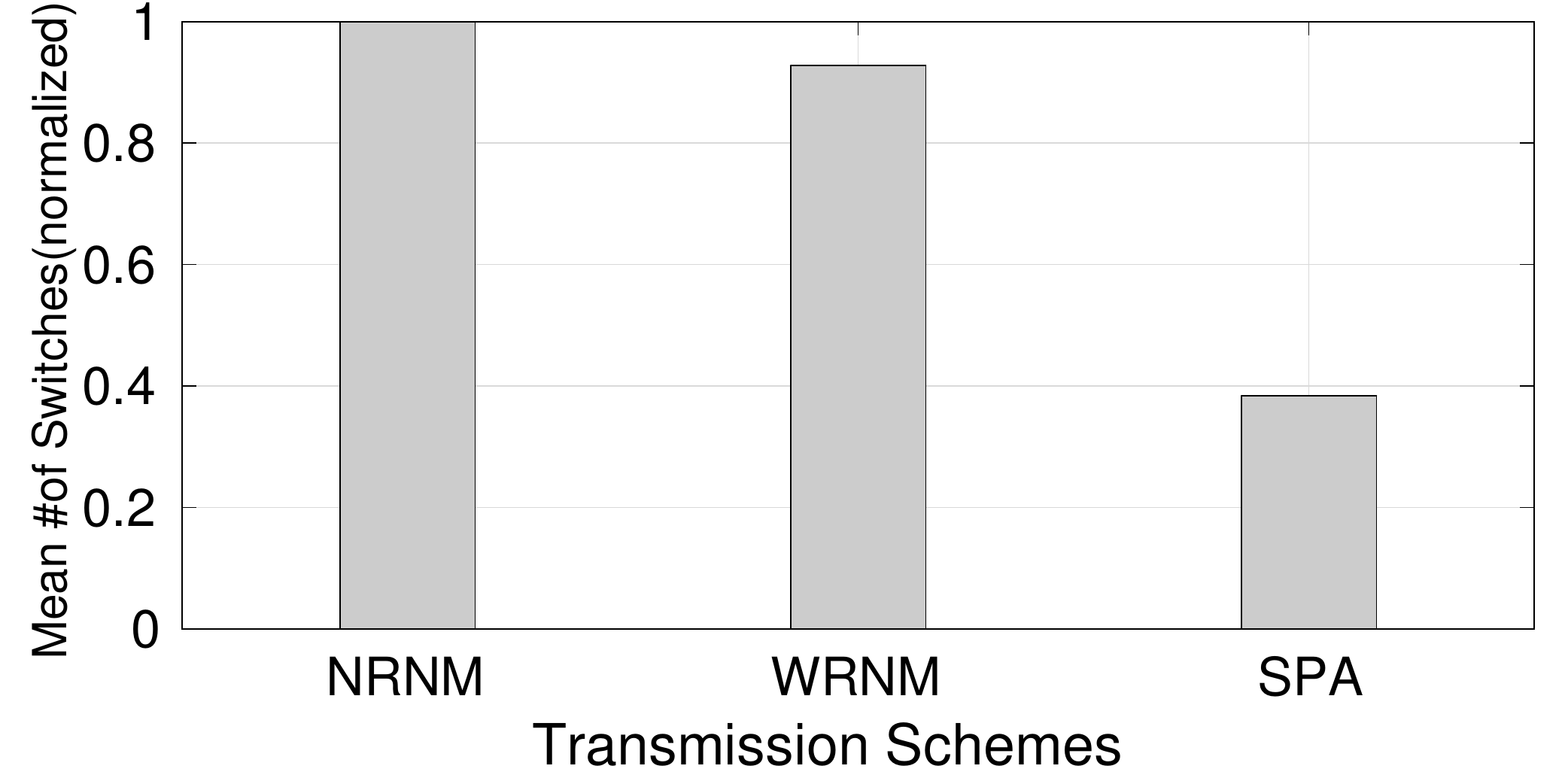}
\label{fig:algovariants_switch}
}
\subfigure[Effect of subset size (r).]
{\includegraphics[width=0.67\columnwidth]{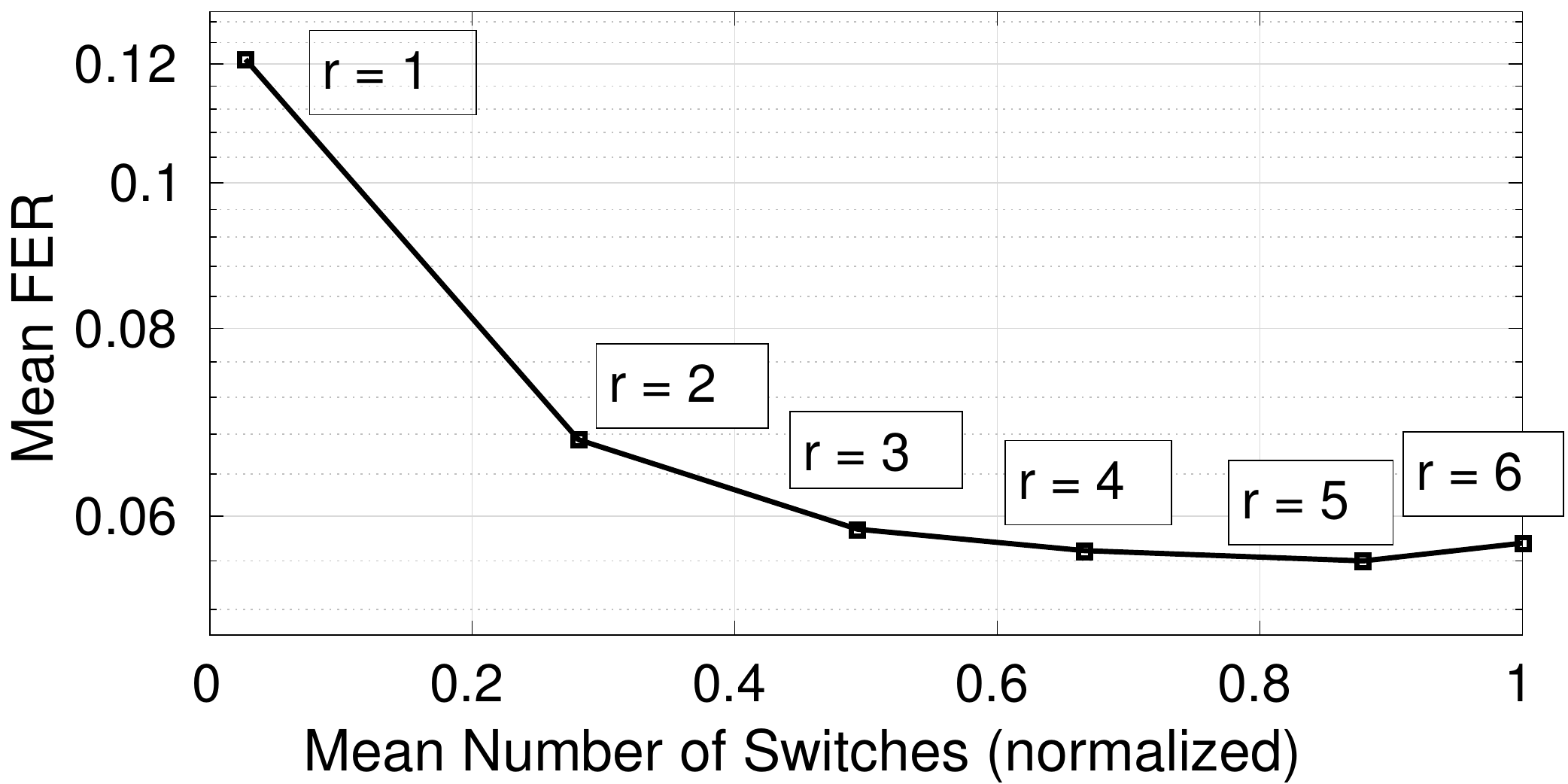}
\label{fig:algo_memory}
}
\caption{Ensemble average performance of different variants of adaptive mode-selection algorithms. }
\label{fig:algo_variants}
\vspace{-1em}
\end{figure*}
\subsection{Evaluating the design of SPA}\label{subsec:expt_performance_adaptive}
To provide meaningful insights into the ingredients in SPA, we adopt the following evaluation methodology. We perform over-the-air experiments over $3$-relay networks in $10$ different topologies and run $860$ frames in every topology for \emph{all $6$} modes of operation in a time-interleaved fashion. From the experimental data (FERs of the individual modes) for these topologies, we create ``samples'' of time-varying $3$-relay topologies, where each sample comprises $4$ topology transitions and with $172$ frames between transitions. The topology at each transition in a sample is chosen uniformly at random, with repetition, and the specific frames are chosen randomly from the (larger) dataset for that topology. We term a collection of such samples of time varying $3$-relay networks as an \emph{ensemble}. We run each of the algorithms over the samples in such an ensemble, and present \emph{ensemble average} results for FER and switching performance to cover as wide a gamut of time-varying $3$-relay networks as possible.
%, while also providing an element of robustness to the interpretation of the results.
In all the experiments, the relaying operation used was DIQIF. The coding rate, modulation, and the parameters for $SPA$ were the same as in Section~\ref{subsec:results_adaptation}, unless otherwise stated.

%\subsubsection{Algorithms Implemented and Nomenclature}
We compared against the following  set of algorithms.\\
\noindent{\bf Baseline Algorithms.} We implement the following:\\
$\bullet$ \emph{DT}: Non-cooperative strategy, where in Phase~II, the Source simply retransmits to the Destination.\\
$\bullet$ \emph{BRUTE} (Exhaustive Search): At every trigger, the empirical FER of every mode is measured, and a decision on the best mode made after observing a set number of frames from each. No machine-learning is used.\\
$\bullet$ \emph{RandPick}: At every trigger (batchwise FER above a threshold), a \emph{random} mode is selected for operation.\\
$\bullet$ \emph{PWR2}: At every trigger, two modes are picked at random. The empirical FER of each is measured and the better is selected (Power of Two Choices Algorithm).\\
\noindent{\bf Adaptive Algorithms.} The following make use of the machine learning techniques outlined in Section \ref{sec:adaptive_algorithms}:\\
$\bullet$ \emph{NRNM}: At every trigger, learning takes place with (i) No Early Reject (NR), and (ii) No Memory (NM), i.e., learning happens across \emph{all possible} modes.\\
$\bullet$ \emph{WRNM}: At every trigger, learning takes place (i) With Early Reject (WR), and (ii) No Memory (NM)\\
$\bullet$ \emph{SPA}: At every trigger, learning takes place (i) With Early Reject (WR), and (ii) With Sorted Memory (SM), i.e., over a \emph{ranked subset} of modes.\\

\vspace{-2mm}
\noindent{\bf Comparison with baseline schemes.}
\begin{figure}
  \centering
  \includegraphics[width=0.95\columnwidth]{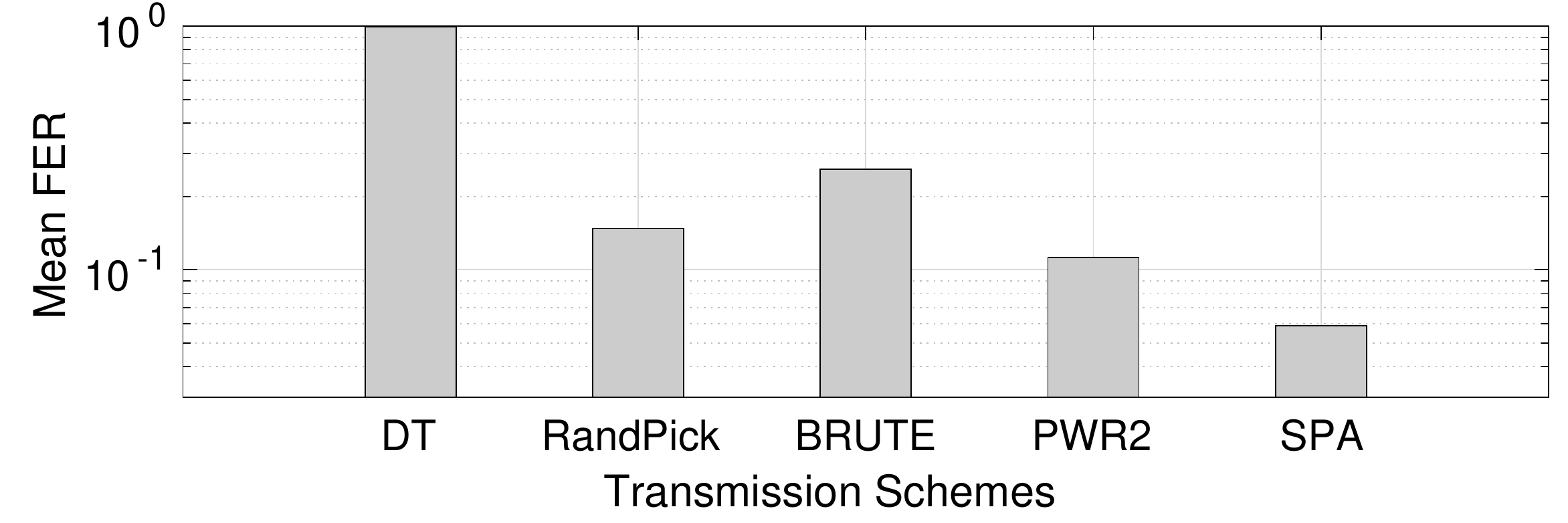}
    \vspace{-1em}
    \caption{Comparison with Baseline Schemes}
  \label{fig:adaptive_baseline}
  \vspace{-1em}
\end{figure}
Fig. \ref{fig:adaptive_baseline} demonstrates the ensemble average FER benefits of SPA over certain baseline strategies in $3$-relay networks with multiple topology changes over time. In our topologies, the direct link was very weak, and it is hence another clear demonstration of the need for using physical layer cooperation to ensure connectivity. As we can see, SPA provides an FER that is $50\%$ of the FER achieved by the best baseline strategy, which in our case, is \emph{PWR2}. {We would expect this gain to be even higher in scenarios (`samples' from our ensemble) where there is a greater variability in the FER across the modes, and choosing bad mode(s) can significantly affect performance. The ensemble averaging technique used to present the results here, inherently ``smooths over'' the sample variability of \emph{PWR2} or \emph{Randpick}'s performance.}\\

\vspace{-2mm}
\noindent{\bf Early rejection and learning over subsets.}
Fig. \ref{fig:algo_variants} depicts ensemble average FER and switching performance for different variants of our learning-based algorithms for adaptive mode switching.
Fig. \ref{fig:algovariants_fer} demonstrates that invoking the early-reject criterion and not allocating equal learning resources to all modes, provides a significant benefit in performance: the NRNM scheme suffers a factor of $2.5$ penalty in average FER over WRNM and SPA, which employ the early reject criterion. This is because the early-reject allows a faster convergence to the requisite mode, while not sending too many frames via the obviously bad modes during the learning process.
Fig. \ref{fig:algovariants_fer} also demonstrates that SPA, which learn over subsets (of size $3$) as opposed to WRNM (which learns across all modes) only suffers a very nominal penalty (less than $0.4\%$) in FER performance. This is because convergence speed is much faster over smaller subsets, and even if a selected mode does not satisfy the FER requirements, the trigger mechanism allows the algorithm to quickly learn from the other partition of modes to converge to the best. In addition, a significant benefit of learning over subsets is observed in the average number of mode switches that are required in the course of running an algorithm over a time-varying $3$-relay network. From Fig. \ref{fig:algovariants_switch}, we can see that the mode-switching overhead reduces by a factor of $\approx 3$ when learning takes place over subsets, again owing to faster algorithmic convergence over smaller sets. Reducing switching overhead is key to network operation, as this means less feedback requirements, less need of (re)-synchronization across terminals, and better network resource utilization in the sense of allocating free relays to other source-destination pairs.\\

\vspace{-2mm}
\noindent{\bf Effect of subset size (memory) on learning. }\label{subsec:effect_of_memory.}
Fig. \ref{fig:algo_memory} depicts ensemble average FER versus switching performance with varying memory (or subset) sizes for learning. For $3$-relay networks, we have at most $6$ modes of operation, and a memory size of $r$ (ranging from $1$ to $6$) in SPA for learning across the top $r$ ranked modes, every time there is a trigger.
We observe in Fig. \ref{fig:algo_memory} that average FER performance starts to improve with increasing memory size before saturating. The maximum gains are observed when increasing the memory size from $1$ to $2$ (about $44\%$). Switching overhead, on the other hand, increases monotonically and significantly with memory size. While a smaller memory size warrants less mode switches during learning, it also suffers from the penalty of sending too many frames through a badly chosen mode before the algorithm can detect a mode that meets the FER requirements. The tradeoff here is interesting as it seems to suggest that as long as we are operating above a certain memory size ($3$ in the case of our networks), the differences in FER performance are insignificant, while we gain a lot in terms of switching overhead with using a smaller acceptable memory size (factor of $\approx 2$ from size $3$ to size $6$).
\begin{figure*}
  \centering
  \includegraphics[width=0.8\textwidth]{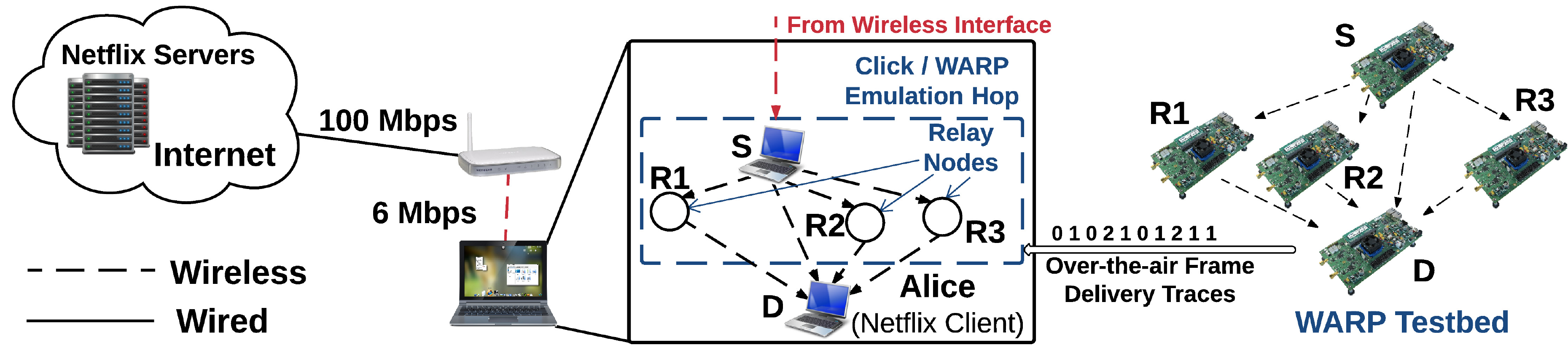}
  \caption{Experiment setup for video streaming.}
  \label{fig:video_exp}
  \vspace{-1em}
\end{figure*}

\section{An Application Perspective:\\Video Streaming}
\label{sec:video_evaluation}
In this section, we evaluate how the benefits of adaptive PHY cooperation translate to video streaming performance and also compare it with adaptive routing. % (as an example for applications that require high data rates).
%To do so,
%We use results from our WARP OTA experiments to emulate PHY cooperation below Netflix.
%
%the benefits of our adaptive physical cooperation for applications running at higher layers of the network stack, for instance, video streaming.
%Currently, video streaming accounts for almost two thirds of all consumer internet traffic \cite{cisco_network_forcast} with speculations of increasing shares in the upcoming years. As a result, and to follow up on our motivating example in the introduction, we construct an experiment to answer : How does Alice's streaming experience improve when using Adaptive Physical Layer cooperation ?
\begin{figure*}
  \centering
\subfigure[Benefits of Adaptation]
  {\includegraphics[width=0.68\columnwidth]{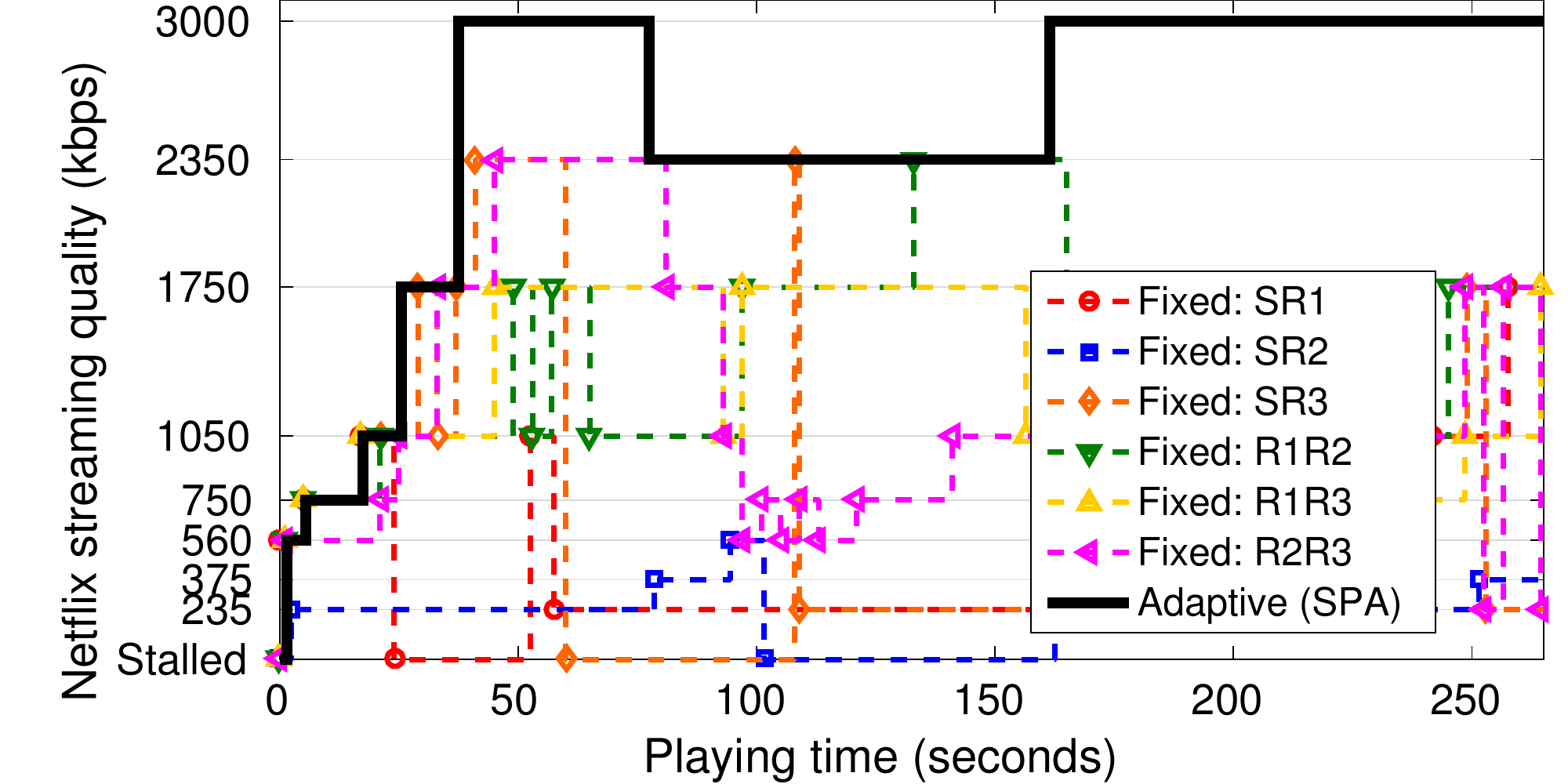}
\label{fig:video_perf_adap_time}
}
\subfigure[Throughput Benefits]
  {\includegraphics[width=0.63\columnwidth]{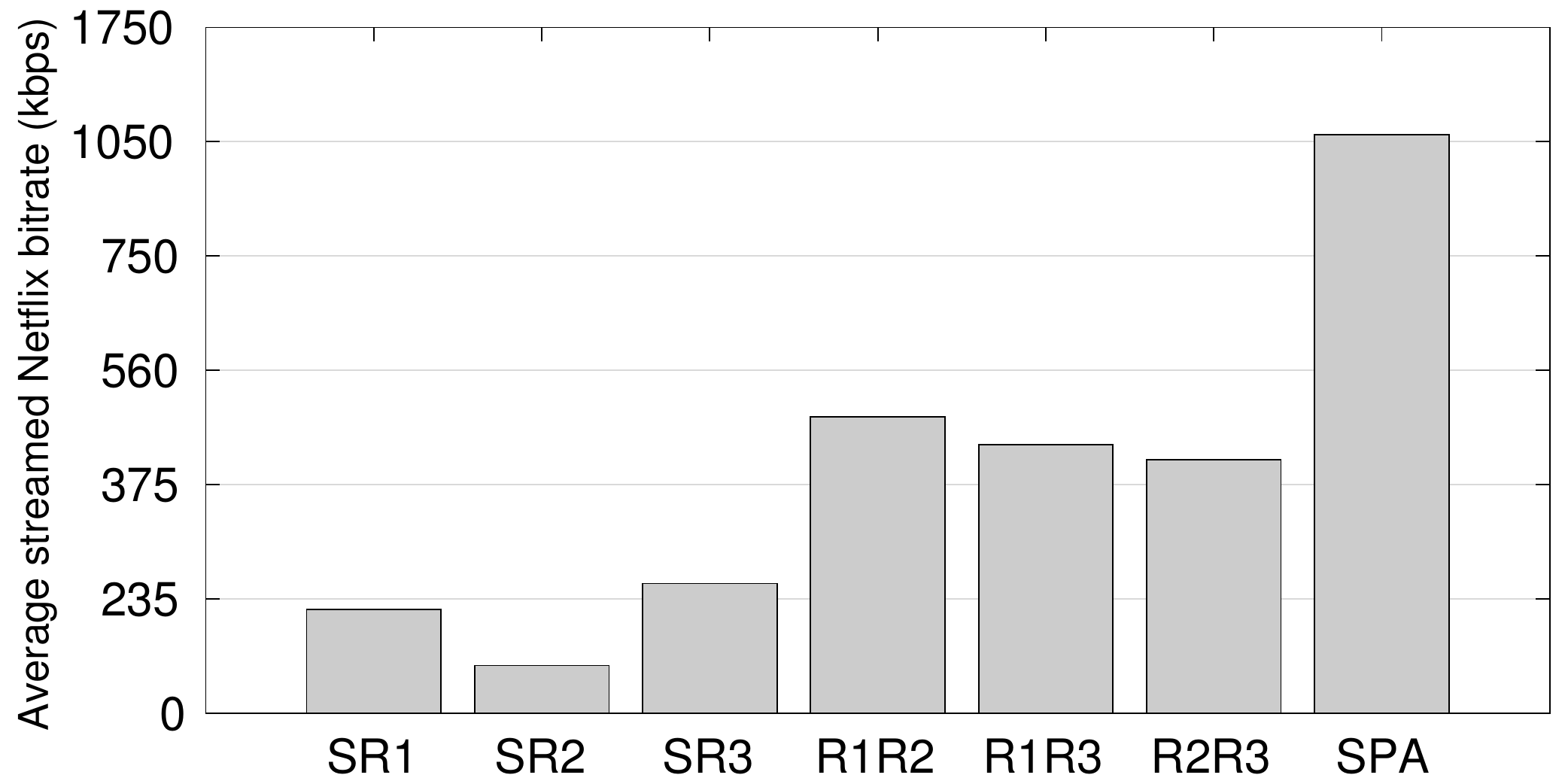}
\label{fig:video_perf_adap_avg}
}
\subfigure[Adaptive Cooperation vs Genie-Routing]{
  \includegraphics[width=0.68\columnwidth]{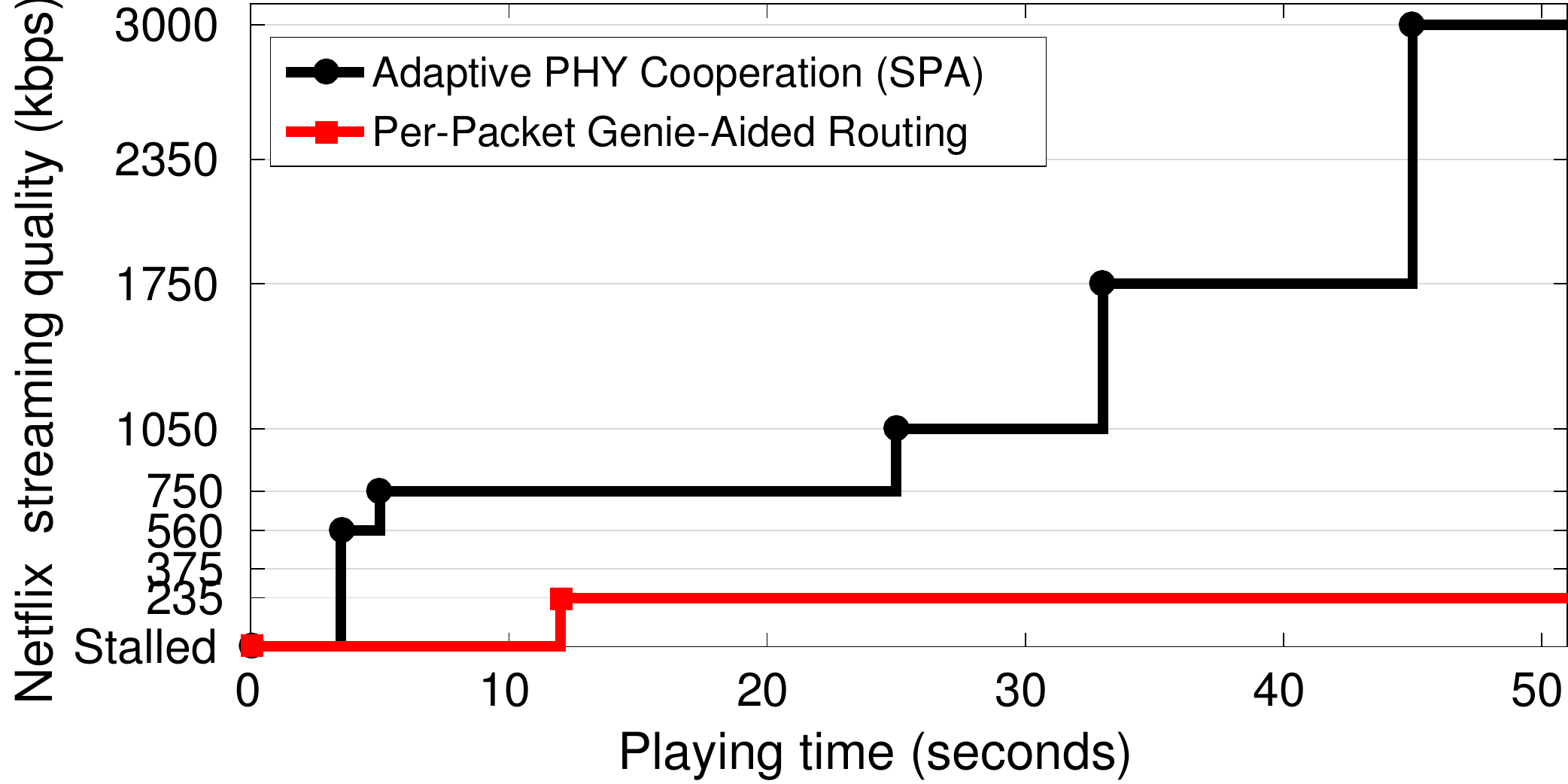}
\label{fig:video_perf_adap_routing}
}
\caption{Video streaming performance of PHY cooperation with SPA.}
\label{fig:video_performance}
  \vspace{-1em}
\end{figure*}
\subsection{Experiment Setup}
We implement the setup in Fig.~\ref{fig:video_exp}: Alice runs a Netflix web browser client on her device (in our experiments, a Linux PC) using an 802.11g wireless router with bit-rate of $6$ Mbps\footnote{The bit-rate is limited so that degradation in performance can directly be observed in video streaming. Recommended Netflix bit-rate for HD streaming is $5$ Mbps.} ; we let the Netflix client stream video from Netflix servers. In between, we introduce an additional hop, that uses PHY cooperation/SPA, emulated using the Click Modular Router toolkit \cite{click}. The Click toolkit runs as a kernel module on the client node. The kernel module intercepts 802.11 frames arriving from the network interface card, filters out frames from the Netflix servers, and passes them through our emulation link before forwarding them to the higher layers. To other applications on the client, Click is a transparent layer in the network stack.\\

%\noindent{\bf Why Emulation?} Ideally, Netflix should be streamed on top of the WARP Testbed, however, we opt to operate the two separately for the following considerations:
%
%\noindent (1) As described in Section \ref{sec:testbed}, the WARP Testbed uses MATLAB for sending and processing frames. Although, this provides flexibility in implementation and testing, it introduces a processing bottleneck (even for some currently standardized PHY layer techniques). Hence, testing Netflix using the Testbed, would introduce slowdowns in the Transport layer that only reflect the processing limits of the MATLAB implementation.\\
%(2) To faithfully compare the gains in video streaming using PHY cooperation/SPA vs. routing, it is necessary to be able recreate similar channel conditions for both tests. To facilitate this, we use the Testbed to multiplex frames that use PHY cooperation and routing. As a result, we cannot run Netflix on top such multiplexing.
%(3) As we see in the next subsection, having a record of the full frame traces highlight the benefits of PHY cooperation/SPA vs. routing by comparing it with a genie-aided routing scheme that has future knowledge of the frame drops and retransmission requests.\\

\vspace{-2mm}
{\noindent{\bf PHY Emulation (Testbed).}
The WARP testbed, described in Section \ref{sec:testbed}, is used to send over-the-air (OTA) frames over a network with three relays assisting a source destination pair. SPA is used to adapt to the best cooperation mode to use. The Testbed generates traces describing how each frame was delivered:\\
(0) Frame received successfully using direct transmission from source to destination.\\
(1) Frame received successfully using selected PHY cooperation mode after failure of the direct transmission.\\
(2) Frame not received successfully using direct transmission and PHY cooperation.

%For frames that use routing only, the Testbed returns:\\
%(0) The frame delivered successfully using the route.\\
%(1) Frame failed delivery from the source to the relay.\\
%(2) a failure in delivering the frame from relay to the destination after successful reception at the relay.

The Testbed uses a payload of 7776 bits corresponding to air times of 180 $\mu$secs and 192 $\mu$secs for the direct transmission and PHY cooperation, respectively.}\\

\noindent{\bf MAC-layer.}
The network interface card for Alice is set to a maximum transmission unit (MTU) of 780, to be in accordance with our WARP experiments MTU and to avoid additional overhead in TCP that can arise due to dropping fragmented frames. {We verified, through TCP flags, that Netflix adjusts to this MTU configuration using Path MTU Discovery.} Our MAC layer retransmission policy is similar to 802.11. A transmission is considered in error (requiring retransmission) if the frame was received incorrectly over both two timeslots (direct and cooperation transmissions). In this case, a retransmission is requested (which consequentially incurs more delay). The maximum number of MAC retransmission is 2. After two retries, the frame is dropped.\\

\noindent{\bf Click.}  We use Click to implement the MAC-layer policy described above. The OTA frame-by-frame traces collected from the WARP Testbed are used by Click to introduce delay(s) and drops to Netflix frames, as would have been experienced by the MAC Layer at Alice when using the Testbed for the PHY cooperation hop: \\
{\it Delay {1} - No Error}: If the frame was correctly received in the first timeslot's direct transmission. {It incurs a transmission delay equal to a single direct transmission.}\\
{\it Delay {2} - No Error}: If the frame was not correctly received in the first timeslot, but was received correctly through cooperation in the second. {The delay added is equivalent to direct and PHY cooperation airtimes.}\\
{\it Delay {2} - Error}: If the frame was received incorrectly through both the direct transmission in the first timeslot, and cooperation in the second timeslot.\\
If a Netflix frame is assigned an error then Click applies the required delays and then views the next frame trace; if again it experiences an error, then that frame is dropped by Click. Otherwise, the corresponding delay is applied and then the frame is forwarded to the higher layer in the device.
We make the assumption, which is valid for our indoor scenario, that the propagation delay is negligible in comparison to the transmission time. The emulated link is modeled {similar} to an ad-hoc 802.11 mode, with disabled RTS and CTS.

Since our delay is emulated, delay jitter is a concern. To meet tight timing delays, we time our delays with spinlocks on a dedicated processor in the node. As a result, in 99.5\% of the frames, the jitter was zero (rounded to the nearest $\mu$ secs) and only $0.01\%$ of the frames suffered from inaccuracies beyond $5\%$ of the target delay.\\

\noindent{\bf Why Emulation?} (1) With the objective of comparing our scheme to adaptive routing, we decided to equip adaptive routing with clairvoyance-i.e., genie-aided knowledge of \emph{future} frame drops and retransmission requests, while providing \emph{no clairvoyance} to PHY cooperation/SPA. This way, we provide our competitor with an unfair advantage that we cannot otherwise provide with a real time implementation. Restricting ourselves to a comparison with any \emph{specific} routing protocol would make the case for PHY cooperation/SPA weaker. (2) To faithfully compare the video streaming gains using PHY cooperation/SPA vs adaptive routing, it is necessary to be able to create similar channel conditions for both tests. This is done by time-multiplexing frames that use PHY cooperation and routing in the testbed. A real time experiment would need to run schemes sequentially, and hence it is not feasible to maintain identical channels conditions over long experiments (of the order of tens of seconds). (3) A frame-by-frame OTA trace (from the testbed) of the PHY FERs is all that is needed by the MAC and higher layers, owing to the layered structure of networks. While giving us the advantages mentioned above, we are able to provide a fair comparison across all schemes in this manner.
\subsection{Video Streaming Performance}
%We here examine what benefits SPA can provide for video streaming, using the setup we described
%To demonstrate the benefits of adaptive physical layer cooperation in a video streaming application, we use the implementation
%in Section~\ref{sec:video_evaluation}.
We performed {long} WARP OTA experiments over multiple $3$-relay configurations (running all $6$ {modes} per configuration). These configurations  create a time-varying sample of $3$-relay networks (with a set number of topology transitions across time) over which we emulate video streaming.
Frame-by-frame traces from these OTA experiments are used by click to implement the MAC-layer with delays, retransmissions and drops.\\
%We compare: (i) The streaming bitrate (from Netflix) with any {fixed} cooperative mode across time, (ii) The streaming bitrate with  SPA, and (iii) The streaming bitrate with per-packet genie-aided routing, i.
%e., when we select the best $S-R_i-D$ link from among $3$ (as a line network), for each transmitted frame. Our results are summarized in Fig.~\ref{fig:video_performance}. In the following, we provide a detailed interpretation of the same.

\noindent{\bf Adaptation and cooperation.}
%===================================================================
Fig.~\ref{fig:video_perf_adap_time} plots the bitrates streamed from Netflix, when the Click emulated last hop uses the traces from the $6$ fixed modes, as well as SPA. There were $20$ different $3$-relay topologies in the sample used to generate the frame-by-frame traces, with a total of $132000$ frames, each.
We see in Fig.~\ref{fig:video_perf_adap_time} that SPA is able to maintain the top $2$ quality levels ($3000$ Kbps and $2350$ Kbps) most of the time, while each  of the $6$ fixed modes suffers a significant hit in streaming rates. Moreover, the performance gap between the fixed modes and SPA increases as time progresses, because, %by offering consistent performance,
%SPA avoids
once the video rate drops due to a momentarily poor mode connection, it does not instantly get restored when the physical channels recover for two reasons:
(1) the slow-start algorithm used by TCP restarts when frames are dropped, and (2) the adaptation strategy in Netflix gradually progresses to higher rates, as opposed to making abrupt jumps. SPA avoids these by offering  consistent performance.
Fig.~\ref{fig:video_perf_adap_avg} plots average streaming bitrates, and shows that SPA provides over $95\%$ average video throughput improvement over the best of the six fixed modes (The non-cooperative direct-transmission (DT) scheme is not plotted as it was mostly in  ``buffer empty''  stalling state).\\
% supported by each of the fixed modes, and by SPA. We see that SPA provides over 95 \% average video throughput improvement over the best fixed mode. Also, while we did not plot the throughput from the non-cooperative direct-transmission (DT) scheme, we noted that its performance was extremely poor: it was mostly in the ``buffer empty'' state, i.e., there was frequent stalling in the video; when not in a stalled state, the best it could achieve was the minimum streaming quality of Netflix, i.e., $235$ Kbps. This clearly demonstrates the potential of physical layer cooperation in facilitating demanding applications that may otherwise be infeasible due to, among other things, a weak WiFi signal over a direct link.

\noindent{\bf Adaptive PHY beats adaptive routing.}
%======================================
One can ask: why not just use an adaptive IP-level routing protocol, that simply finds the best path to connect $S$ and $D$, and uses the relays for routing as opposed to PHY layer cooperation?  To answer this, we put our adaptive PHY cooperation strategy to the test, against a per-packet genie-aided routing protocol that we implement over our Click-emulated link.
The genie-aided protocol selects the best possible route, given advance knowledge of {\em all future frame traces}, and thus outperforms every implementable protocol.
In particular: \\
$\bullet$ Each link allows up to $4$ max  MAC retransmissions.\\
$\bullet$ If a packet can be delivered to $D$ through any path, it selects the path that requires the least total number of retransmissions, and thus deliver the packet faster.\\
$\bullet$ If the packet cannot be delivered - given the maximum retransmissions at each link - send the packet along the route that drops it first, and thus, incurs the least delay.
Table \ref{tab:genie_routing} also shows the reduction in packet drop ratio when using SPA over the genie-aided protocol.

\begin{table}[t!]
    \small
    \centering
\begin{tabular}{|l|c|}
\hline
{\bf Strategy} & {\bf Packet drop rate} \\ \hline
$S-R_1-D$ & 27.63 \%\\ \hline
$S-R_2-D$ & 13.70 \% \\ \hline
$S-R_3-D$ & 11.44\% \\ \hline
genie-aided routing & 1.60 \% \\ \hline
SPA + PHY Cooperation & 0.023 \% \\ \hline
\end{tabular}
\caption{Packet drop rates for video streaming.}
\label{tab:genie_routing}
\vspace{-1em}
\end{table}

Fig. \ref{fig:video_perf_adap_routing} plots the bitrates streamed from Netflix (for a $50$ sec video play time) with  genie-aided routing and SPA.
%, and the physical layer cooperation with SPA in the emulated module, as described above. We observe from the plots that the
Genie-aided routing makes a very slow start and is at most able to support the lowest video quality; in contrast,  SPA %(adapting across only $3$ $2$-relay modes, due to the killed direct S-D link)
allows Netflix to quickly start streaming and gradually improve the quality of video streamed, reaching the highest quality of $3000$ kbps towards the end of the play time, achieving an overall $6\times$ throughput gain.
\iffalse
Genie-aided routing is only able to support up to $1050$ kbps streaming rate in the beginning and is soon overwhelmed by packet drops and delays and buffering for more than half of the play time; in contrast,  SPA %(adapting across only $3$ $2$-relay modes, due to the killed direct S-D link)
allows Netflix to quickly start streaming and gradually improve the quality of video streamed, reaching the highest quality of $3000$ kbps and maintain it until the end of the play time.
\fi
\iffalse
Genie-aided routing makes a very slow start and is at most able to support the lowest video quality; in contrast,  SPA %(adapting across only $3$ $2$-relay modes, due to the killed direct S-D link)
allows Netflix to quickly start streaming and gradually improve the quality of video streamed, reaching the highest quality of $3000$ kbps towards the end of the play time, achieving an overall $6\times$ throughput gain.
\fi
This can be explained by the fact that routing along a path requires the relay to perfectly decode; SPA requires only the receiver to do so, leading to lower rates of frame delivery failures and consequentially less frame drops. The genie-aided routing protocol suffers from higher frame drops, which causes TCP/IP to misinterpret it as network congestion and therefore degrade its end-to-end throughput.
%. The average throughput gain of physical layer cooperation with SPA over genie-aided routing is upwards of 6x.

\section{Conclusions}
\label{sec:conclusions}
We presented the design and experimental evaluation of adaptively selecting a small subset of relays to assist an S-D communication via PHY cooperation. Our main conclusions are:
(1) The returns from using multiple relays for PHY cooperation diminishes progressively; a well-chosen (small) subset extracts a large portion of the potential gains, while keeping complexity and overhead demands reasonable.
%(1) PHY cooperation with $2$ relays can sometimes (but not always) outperform PHY cooperation with $1$ relay, and is thus worth considering. Yet, we also noted a trend of diminishing returns with an increasing number of active relays in the network.
(2) For practical PHY cooperation, using network channels to select the highest capacity subnetwork is not a good choice (unlike for routing).
(3) We get significant benefits, and a consistent performance across topology changes, by having SPA adaptively select which sub-network to use.
% of relay(s) to use for PHY cooperation
%. SPA is an algorithm that enables such gains at low complexity. SPA is conceptually simple: it selects modes via FER training; yet, it required detailed and careful design in order to learn the best mode with low overhead.\\
%(3) PHY cooperation with SPA can enable consistently high rates when streaming Netflix--upto 2x throughput gain over using the best non-adaptive mode were observed in our experiments.\\
(4) PHY cooperation with SPA shows a 6x gain in Netflix streaming throughput, over genie-aided best-path routing (that does not provide PHY cooperation).

\appendix
\section{Outage Probability Derivation}\label{sec:app_outage}
\newcommand{\om}{\Omega}
\newcommand{\omc}{{\Omega^c}}
\newcommand{\lam}{\lambda}
\newcommand{\myprob}[1]{\mathbf{Pr}\left\{ #1 \right\}}
The approximate capacity of the $N$ relay diamond network \emph{with a direct link} (within an additive constant) is the minimum over $2^N$ mutual information terms.
\[
\bar{C} \approx \min_{\om} I(X_\om; Y_\omc | X_\omc)
\]
Each cut $\om \subseteq \{1,2,\ldots,N\}$, and the mutual information for the corresponding cut is denoted by $I_\om$ from here on. As shown in \cite{Avestimehr2015,avestimehr_hd}, $I_\om$ can be approximated as follows in terms of the capacity of the $2N+1$ links in the network:

{\small
\begin{align}
I_\om \approx \max \left\{\begin{array}{c}\log(1+h_{sd}^2),\\ \left\{\max\limits_{i\in \om} \{ \log (1+h_i^2)\} + \max\limits_{j\in \omc}\{ \log (1+g_j^2)\}\right\} \end{array}\right\}
\label{eq:cut_maxexpr}
\end{align}
}

where $h_{sd}$, $h_i$ and $g_j$ denote the \emph{absolute values} of the $S-D$, $S-R_i$ and $R_j-D$ channels. The signal and noise powers are normalized to unity.

For a Rayleigh fading channel model, $h_{sd}^2 \sim \mathbf{Exp}(\lam)$, $h_i^2 \sim \mathbf{Exp}(\lam_i^l)$ and $g_i^2 \sim \mathbf{Exp}(\lam_i^r)$.  Given a rate $R$, the probability that $\bar{C}$ is less than $R$, i.e., the outage probability, can be routinely upper bounded as:

{
\small
\begin{align*}
P_{\text{out}}(R,h_{sd},\{h_i,g_i\}_{\{i \in [1:N]\}}) &= \mathbf{Pr}\{\bigcup_\om I_\om < R\}
                                                       \leq \sum_\om \myprob{I_\om < R}
\end{align*}
}
In practice, the upper bound is fairly tight. From \eqref{eq:cut_maxexpr},

{
\small
\begin{align*}
    \myprob{I_\om < R}\stackrel{(a)}{\approx}& \begin{array}{l}\myprob{\log(1+h_{sd}^2) < R}\\ \times \myprob{ \begin{array}{l} \log (1+\max_{i\in \om} \{h_i^2\}) \\+ \log (1+\max_{j\in \omc}\{g_j^2\})\end{array} < R}\end{array}
\end{align*}
}
where $(a)$ is true because $h_{sd}$ is independent of the $h_i$'s and $g_j$'s. Since $h^2 \sim \mathbf{Exp}(\lam)$, we have
\begin{equation*}
\myprob{\log(1+h_{sd}^2) < R} = 1- e^{-\lam (2^R-1)}
\end{equation*}
For the second term in the product, $X \triangleq \max_{i\in \om} \{h_i^2\}$ and $Y \triangleq \max_{j\in \omc}\{ g_j^2\}$. Then the second term becomes
\[
P_\om = \myprob{\log(1+X)+\log(1+Y)<R}
\]
If $F_X(x)$ is c.d.f of $X$ and $f_X(x)$ is its p.d.f,
\[
P_\om = \int_{x=0}^{2^R-1} f_X(x)F_Y\left(\frac{2^R}{1+x}-1\right) dx
\]
Now, $F_X(x)$ and $f_X(x)$ (similarly $F_Y(y)$, $f_Y(y)$) are:
\begin{align*}
&F_X(x) \stackrel{(b)}{=}\prod_{i\in \om} \myprob{h_i^2\leq x} = \prod_{i\in \om} (1-e^{-\lambda^l_ix})\\
&f_X(x) = \frac{d}{dx} F_X(x) = \sum_{i\in \om} \lambda_i^l e^{-\lambda_i^lx} \prod_{k\in \om\setminus\{i\}} (1-e^{-\lambda^l_kx})
\end{align*}
where $(b)$ is true since $h_i$'s are independent.
$P_\om$ can now be written as a (weighted) sum of a constant and several integrals as $P_\om^t = \int_{x=0}^{2^R-1} e^{-\alpha x} e^{-\beta (\frac{2^R}{1+x}-1)} dx$. Then,
\[
    P_{\text{out}}(R,h_{sd},\{h_i,g_i\}_{\{i \in [1:N]\}})\approx (1- e^{-\lam (2^R-1)}) \sum_\om P_\om
\]
For numerical evaluation, we essentially need to \emph{compute} terms of the type $P_\om^t$ above.\\
\noindent\emph{Optimization problem:} Given $N$ relays, we want to select
an \emph{outage optimal} subnetwork with $k$ relays $(k < N)$. To do
so, we need to solve the following problem:
\begin{align}
  P_{\text{out}}^{\text{opt}}(R,k) =
  \arg\min_{\mathcal{I}:|\mathcal{I}|=k}
  P_{\mathsf{out}}(R,h_{sd},\{h_i,g_i\}_{\{i \in
    \mathcal{I}\}})\label{eq:optimization_prob_outage}
\end{align}

\bibliographystyle{acm}
\bibliography{refs}
%\section{SPA and LEARN Parameters}\label{sec:app_SPA_LEARN_parameters}
%\input{appendix_SPA_LEARN_parameters.tex}

\end{document}